\documentclass[twocolumn, times, trackchanges]{aastex63_bren}

\usepackage[ruled,vlined]{algorithm2e}
\usepackage{graphicx}
\usepackage{bbm}
\usepackage{float}
\usepackage{xspace}
\usepackage{listings}
\usepackage[hypertexnames=false]{hyperref}
\usepackage{subfiles} 
\usepackage[encapsulated]{CJK}

\hypersetup{linkcolor=red,citecolor=green,filecolor=cyan,urlcolor=magenta}

\defcitealias{baron19}{B19}
\defcitealias{baron24}{B24}

\newcommand{\halpha}{H$\alpha$\xspace}
\newcommand{\hbeta}{H$\beta$\xspace}

\newcommand{\oiii}{$\text{[O\,{\sc iii}]}$\xspace}

\newcommand{\nii}{$\text{[N\,{\sc ii}]}$\xspace}
\newcommand{\siifull}{$\text{[S\,{\sc ii}]}\lambda\lambda \, 6717\mathrm{\AA}+6731\mathrm{\AA}$\xspace}
\newcommand{\sii}{$\text{[S\,{\sc ii}]}$\xspace}

\newcommand{\oiiihbeta}{$\log (\text{[O\,{\sc iii}]}/\text{H}\beta)$\xspace}
\newcommand{\niihalpha}{$\log (\text{[N\,{\sc ii}]}/\text{H}\alpha)$\xspace}
\newcommand{\siihalpha}{$\log (\text{[S\,{\sc ii}]}/\text{H}\alpha)$\xspace}
\newcommand{\oihalpha}{$\log (\text{[O\,{\sc i}]}/\text{H}\alpha)$\xspace}

\newcommand{\mic}{$\mathrm{\mu m}$\xspace}

\shorttitle{starburst to post-starburst survey}
\shortauthors{Baron et al.}

\begin{document}

\title{Sparks: The Magellan/FIRE survey from starburst to post-starburst}

\author[0000-0003-4974-3481]{Dalya Baron}
\email{dalyabaron@gmail.com}
\affiliation{Kavli Institute for Particle Astrophysics \& Cosmology, Stanford University, CA 94305, USA}
\affiliation{Center for Decoding the Universe, Stanford University, CA 94305, USA}

\author[0000-0003-4075-7393]{David J. Setton}\thanks{Brinson Prize Fellow}
\affiliation{Department of Astrophysical Sciences, Princeton University, 4 Ivy Lane, Princeton, NJ 08544, USA}

\author[0000-0002-0463-9528]{Yilun Ma (\begin{CJK*}{UTF8}{gbsn}马逸伦\ignorespacesafterend\end{CJK*})}
\affiliation{Department of Astrophysical Sciences, Princeton University, 4 Ivy Lane, Princeton, NJ 08544, USA}

\author[0000-0002-7738-6875]{J.~X.~Prochaska}
\affiliation{Department of Astronomy and Astrophysics, University of California Santa Cruz, 1156 High Street, Santa Cruz, CA 95064, USA}
\affiliation{Kavli Institute for the Physics and Mathematics of the Universe (Kavli IPMU), 5-1-5 Kashiwanoha, Kashiwa, 277-8583, Japan}
\affiliation{Division of Science, National Astronomical Observatory of Japan, 2-21-1 Osawa, Mitaka, Tokyo 181-8588, Japan}

\author[0000-0003-4693-6157]{Gabriela Canalizo}
\affiliation{Department of Physics and Astronomy, University of California, Riverside, 900 University Ave., Riverside CA 92521}

\author[0000-0003-4949-7217]{Ric Davies}
\affiliation{Max-Planck-Institut f\"ur extraterrestrische Physik, Giessenbachstra{\ss}e, 85748 Garching, Germany}

\author[0000-0002-5612-3427]{Jenny~E.~Greene}
\affiliation{Department of Astrophysical Sciences, Princeton University, Princeton, NJ 08544, USA}

\author[0000-0003-0291-9582]{Dieter Lutz}
\affiliation{Max-Planck-Institut f\"ur extraterrestrische Physik, Giessenbachstra{\ss}e, 85748 Garching, Germany}




\begin{abstract}
Rapid transitions from starburst to quiescence constitute a key evolutionary pathway in galaxy formation. Post-starburst galaxies trace this brief phase, exhibiting optical spectra dominated by intermediate-age stellar populations with strong Balmer absorption features. Although rare locally, such systems are commonly revealed by \textit{JWST} observations among massive galaxies at $z \gtrsim 3$. In the nearby Universe, their evolutionary stage remains uncertain: Balmer-strong galaxies hosting active galactic nuclei (AGN) show conflicting star formation rates (SFRs), with optical diagnostics implying quenching while far-infrared emission suggests ongoing obscured star formation. We present \textit{Sparks}, an infrared survey designed to study the transition from starburst to post-starburst. Using the FIRE spectrograph on the Magellan Telescope, \textit{Sparks} provides near-infrared spectra (0.82--2.51~$\mu$m) for 93 local massive galaxies spanning three orders of magnitude in SFR, from starbursts to quenched post-starbursts, including AGN hosts. Here, we describe the survey goals, sample selection, observations, and data reduction, and examine galaxy properties derived from stellar population synthesis fitting of photometric data covering far-ultraviolet to far-infrared. Our new panchromatic-based SFR and star formation history measurements divide the sample into three groups: galaxies undergoing their first major starburst in the past $\sim 1$ Gyr; galaxies undergoing their second major starburst, with optical continua dominated by intermediate-age stellar populations formed during the previous recent burst; and post-burst quenching systems. AGN appear predominantly in the second group, explaining why systems with strong Balmer absorption and AGN show elevated far-infrared emission, and implying a short delay between starburst and black hole accretion.
\end{abstract}

\keywords{Galaxy evolution (594), Post-starburst galaxies (2176), Starburst galaxies (1570), Star formation (1569), Near infrared astronomy (1093)}

\section{Introduction}\label{sec:intro}

The local galaxy population exhibits a clear bimodality in color, with the population divided into blue star-forming spirals and red quiescent ellipticals. This dichotomy is also reflected in their stellar masses, molecular gas contents, and stellar kinematics (e.g., \citealt{strateva01, kauff03b, baldry06, baldry08, bell04, bundy05, saintonge17}). Observations of quiescent galaxies reveal that their number density and stellar mass density increase since $z \sim 3$--4 (e.g., \citealt{bell04, faber07, ilbert13, muzzin13}), suggesting that galaxies transition from the star-forming `blue cloud' to the quiescent `red sequence' over cosmic time, a process called galaxy quenching. Galaxy quenching can range from `slow' to `fast': the former takes place over billions of years and is associated with gas depletion and/or preventive feedback, while the latter takes place over hundreds of millions of years (Myrs) and is triggered by external processes such as mergers or ram-pressure stripping (e.g., \citealt{schawinski14, peng15, maltby18, pawlik18, rowlands18, belli19, wild20, tacchella22}). Constraining the distribution of quenching times is a key goal in galaxy evolution studies (e.g., \citealt{pacifici16, rowlands18, tacchella22}).

Post-starburst galaxies, also known as E+A or Balmer-strong galaxies \citep{dressler83, couch87, poggianti99}, are believed to be a key `fast' transitional phase in the evolution of galaxies from the blue cloud to the red sequence (see review by \citealt{french21}). Their rest-frame optical spectra are characterized by strong Balmer absorption lines or a significant Balmer break, indicating a substantial population of intermediate age A- and F-type stars that outshine the old (age$\sim$1--10 Gyr) stellar populations.  The lack of O- and B-type stars, alongside weak nebular emission\footnote{Or detected line emission that is inconsistent with ionization by young and massive stars, e.g., with line ratios consistent with AGN ionization or shock excitation (e.g., \citealt{alatalo16a})}, has been interpreted as a signature of a starburst that has quenched abruptly (e.g., \citealt{dressler83, couch87, poggianti99, wild07, yesuf14, alatalo16a, french18, pawlik18, baron22}). Their fraction among massive galaxies increases with redshift from $< 1$\% at $z \sim 0$ to $> 5$\% at $z \sim 2$ (\citealt{wild16, rowlands18}). At higher redshifts of $z \gtrapprox 3$, massive quiescent galaxies often show post-starburst signatures (\citealt{belli19, dEugenio20}), though the underlying causes may be different than those of low-$z$ post-starbursts, with numerous recent discoveries owing to the unprecedented sensitivity of the James Webb Space Telescope (JWST; e.g. \citealt{carnall23, carnall23b, dEugenio23, strait23, deGraaff24, setton24, slob24}).

In the local Universe, post-starburst galaxies are suggested to be the evolutionary link between gas-rich major mergers and quiescent ellipticals (e.g., \citealt{zabludoff96, canalizo00, canalizo01, yang04, yang06, kaviraj07, wild09, canalizo13, yesuf14, cales15, wild16, almaini17}). However, recent studies suggest additional processes that lead to post-starburst signatures, such as cyclical evolution within the blue sequence or rejuvenation of red-sequence galaxies \citep{pawlik18}.  In the traditional picture, a major merger of gas-rich spirals triggers a starburst and accretion onto the supermassive black hole (which is visible as an active galactic nucleus; AGN;\citealt{sanders88, mihos94, mihos96, barnes96, springel05, hopkins06}). During the peak of the starburst, the radiation of the stars and AGN are highly obscured by dust, with most of the light emitted in infrared wavelengths. The molecular gas is quickly depleted by the starburst and possibly removed by stellar and AGN feedback, leading to the termination of star formation and black hole accretion. With their star formation quenched, the systems become optically thin, with observed spectra that are dominated by the post-burst A-type population. If the molecular gas has been completely depleted, these systems are then expected to transition into red and dead ellipticals in a few hundreds of Myrs.

Several key aspects of this picture remain uncertain, leading to the following open questions: (I) What is the star formation rate of galaxies transitioning from starburst to quiescence and how can it be robustly estimated using observations at different wavelengths? (II) What is the timing of the starburst–AGN connection? (III) What role does the AGN play in driving the galaxy's transition to quiescence? and (IV) How do the molecular gas properties evolve during this transition, and to what extent is the molecular gas reservoir depleted in post-starburst galaxies?

The star formation properties of post-starburst galaxies remain debated (e.g., \citealt{wild07, pawlik18, rowlands15, french18, smercina18, wild20, baron22, smercina22, suess22, baron23, wu23, wild2025, cenci25, zhu25, setton25}), with optical and infrared tracers sometimes giving conflicting views of their evolutionary stage. \citet{baron22, baron23} reported large discrepancies in SFRs for Balmer-strong galaxies with strong emission lines—especially those hosting AGN—but good agreement for weak-line systems. In the former, ultraviolet–optical continua suggest a burst that ended a few hundred Myr ago, while infrared luminosities ($L_{\mathrm{TIR}}>10^{11}\,L_{\odot}$ in 10--50\% of cases) imply ongoing star formation. For Balmer-strong galaxies with $L_{\mathrm{TIR}}<10^{11}\,L_{\odot}$, \citet{wild2025} found that the relation between the total infrared luminosity ($L_{\mathrm{TIR}}$) and the H$\alpha$ luminosity deviates from that expected under the assumption that dust traces star formation. The deviation increases with decreasing H$\alpha$ equivalent width and particularly evident for EW(H$\alpha$) $< 3\,\mathrm{\AA}$, leading them to suggest that in these systems, dust is more dominated by the ambient interstellar medium and its infrared emission cannot be used to estimate SFR.

Several explanations have been put forward to resolve this discrepancy, including (I) There is significant obscured star formation that does not appear through the stellar continuum emission (\citealt{baron22, baron23}). In this case, optical observations might not trace all of the star formation. (II) The infrared radiation, which has an averaging timescale of $\sim 100$ Myr, may lead to an overestimation of the instantaneous star formation rate in galaxies that experienced a recent dramatic decrease in star formation (e.g., \citealt{hayward14, smercina18, leja19b}). As such systems are expected to still be quite dusty, neither the far infrared nor the optical would trace the star formation rate accurately. Testing this scenario requires observations that trace the star formation rate on short timescales ($\sim 10$ Myr), while being less affected by large dust columns. (III) The infrared radiation is at least partially powered by the central AGN (e.g., \citealt{wu23}). The AGN may potentially affect both optical and infrared tracers, and estimating the star formation rate robustly might therefore require galaxy-AGN decomposition. 

Each of these scenarios has different implications for the starburst-AGN connection. If optical observations trace robustly the star formation in post-starburst galaxies, then they can be used to date the starburst. Such techniques have been applied to post-starburst galaxies with line ratios consistent with AGN ionization, with studies finding a delay of $\sim$200 Myrs between the peak of the starburst and the onset of black hole accretion (\citealt{wild10, yesuf14}). If far-infrared observations robustly trace star formation that is obscured in optical wavelengths, then the starburst-AGN connection has to be revised. In particular, post-starburst galaxies with line ratios consistent with AGN ionization show $L_{\mathrm{TIR}}>10^{11}\,L_{\mathrm{\odot}}$ in 30--45\% of the cases (\citealt{baron22}), suggesting ongoing starbursts. This could therefore suggest that there is a shorter or no delay between the starburst and black hole accretion. Shorter delays between starburst and AGN activity have been inferred by \citet{davies07} using near-infrared integral-field spectroscopy of Seyfert galaxies and by \citep{ellison25} employing machine-learning merger classifications and multi-wavelength AGN diagnostics, both indicating AGN onset within $\sim 50-100$ Myr of the starburst peak.

These scenarios also have different implications to the state of molecular gas in post-starbursts, and its connection to the end of star formation (\citealt{french15, rowlands15, alatalo16b, yesuf17, french18, li19, bezanson22, smercina22, baron23, zhu25, setton25, zanella25}). Using optical observations to derive star formation rates and mm observations to constrain the cold molecular gas mass, studies find unusually massive molecular gas reservoirs (with respect to their star formation rate) in post-starburst galaxies (e.g., \citealt{french15, alatalo16b, baron23}), which contradicts the traditional model where the starburst ends when the molecular gas reservoir is  depleted (e.g., \citealt{barnes96, springel05a, hopkins06}). On the other hand, far-infrared observations suggest that systems with unusually large molecular gas reservoirs are in fact obscured starbursts, implying that there is no tension with the traditional picture that star formation ends when the molecular gas is depleted (\citealt{baron23}). An exception are the galaxies in the SQuIGG$\vec{L}$E sample, where far-infrared observations are not available for their molecular gas-rich post-starbursts, but their H$\alpha$ emission and mid-infrared emission suggests low specific star formation rates \citep{zhu25, setton25}. 

Addressing these questions requires a multi-wavelength approach that captures the full evolutionary sequence, spanning systems in the peak of their starburst to those quenched $> 1$ Gyr ago. At the same time, it is necessary to consider systems with diverse optical emission line properties, including those with gas ionized by young massive stars, AGN, or a combination of both, as well as systems with weak or undetectable lines. Since these systems can be heavily obscured by dust, it is necessary to supplement the rest-frame ultraviolet and optical observations with infrared observations that can peer deeper into their centers, and can be used to identify obscured star formation and black hole accretion.

In this paper we present \emph{Sparks}, an infrared survey of the transition from starburst to post-starburst. The survey uses the FIRE echelle spectrograph on the Magellan Telescope to obtain near-infrared spectra (0.82--2.51 \mic) of a sample of 93 galaxies at different stages in the evolution from starburst to quenched post-starburst. They have diverse emission line properties that include star-forming galaxies and AGN hosts, as well as traditional E+A galaxies with weak emission lines. Together with optical spectra from the Sloan Digital Sky Survey (SDSS; \citealt{york00}), the survey achieves rest-frame spectral coverage of 0.4–2.2 \mic, with two gaps in the infrared due to telluric absorption.

This paper presents the survey goals, sample selection, observations and data reduction, and general galaxy properties derived from stellar population synthesis  (SPS) fitting. In a companion paper (hereafter Paper II; Baron et al. submitted), we present the collection of photometry spanning the far-ultraviolet to the far-infrared for the \textit{Sparks} galaxies. We fit the multi-wavelength SEDs, as well as the optical spectra, using different codes (\texttt{Prospector}; \citealt{leja17}; and \texttt{MAGPHYS}; \citealt{daCunha08}) under a range of model assumptions. Paper II compares the resulting galaxy properties—such as stellar mass and star formation rate—across input data (spectroscopy versus photometry), fitting codes, and modeling choices, and identifies the set of adopted galaxy properties used in this work.

The paper is organized as follows. Section~\ref{sec:sample} describes the sample selection and general galaxy properties. Section~\ref{sec:fire} details the FIRE observations, data reduction, and analysis procedures. In Section~\ref{sec:results}, we use the new, carefully benchmarked galaxy properties from Paper II to refine the position of the Sparks galaxies within the starburst to post-starburst sequence. Section~\ref{sec:science_goals} outlines the main science goals of the program and shows representative FIRE spectra and infrared line detection statistics to illustrate its scientific potential. We summarize in Section~\ref{sec:summary}.

Throughout this paper, we use the term starburst-to-post-starburst sequence to describe the temporal transition an individual galaxy may undergo. The \textit{Sparks} sample is not meant to define a single evolutionary track; rather, it includes galaxies occupying different regions and, perhaps, different classes, of this transition. There is growing evidence that galaxies follow multiple pathways from starburst to quiescence (e.g., \citealt{pawlik18}), and our sample should reflect this diversity.

\begin{figure*}
	\centering
\includegraphics[width=1\textwidth]{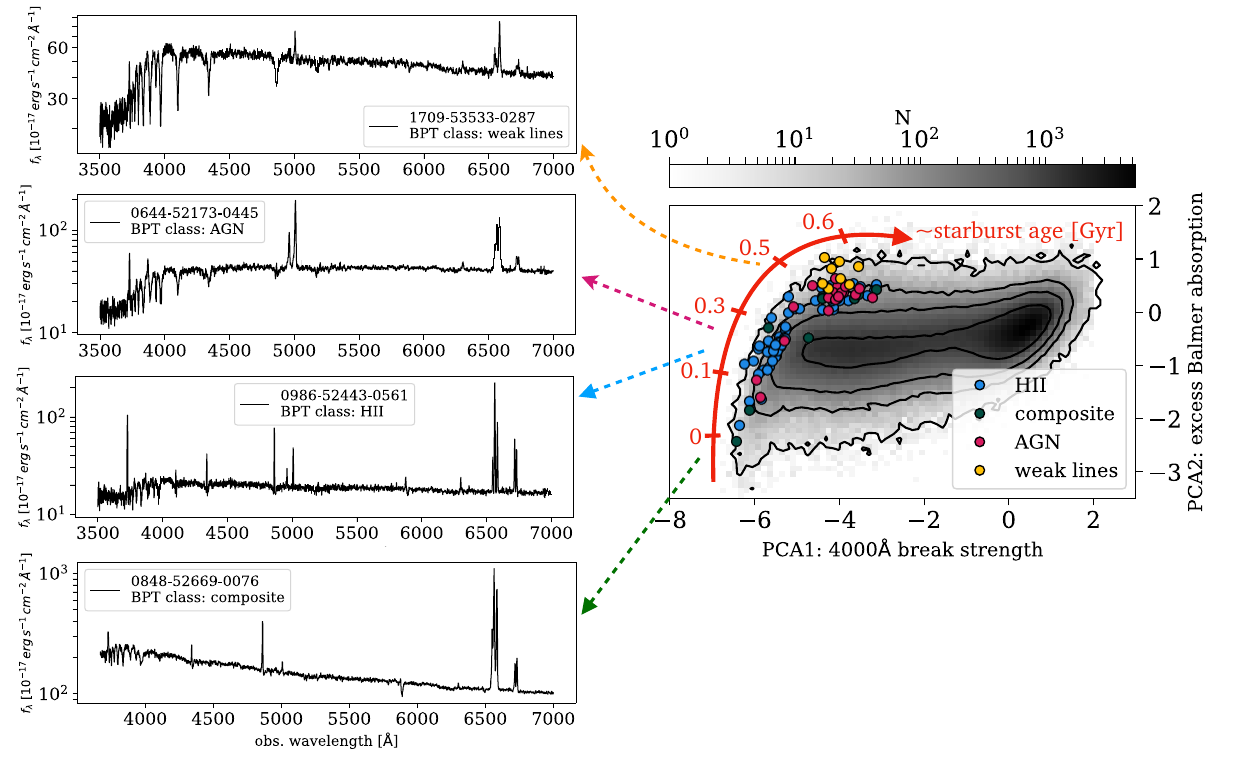}
\caption{\textbf{The starburst to post-starburst sequence used to define our sample.} The right panel demonstrates the use of the 4000~$\mathrm{\AA}$ break and Balmer absorption strengths in defining galaxies along the starburst to post-starburst sequence. The black and white colors represent the two-dimensional distribution of $\sim$650\,000 local ($0.02 < z < 0.2$) galaxies from SDSS DR7 in the two-dimensional {\sc pca} plane defined by \citet{wild07}. This space is based on a PCA decomposition of spectra around the 4000~$\mathrm{\AA}$ break, with PCA1 correlating with the 4000~$\mathrm{\AA}$ break strength and PCA2 with the excess Balmer absorption for a given 4000~$\mathrm{\AA}$ break strength. The starburst to post-starburst sequence as identified by \citet{wild07} occupies the left tail of the two-dimensional distribution, where starbursts are found in the bottom left, and as the starburst ages, the system moves roughly along the red track. The starburst ages indicated along the track are adopted from \citet{wild10}, who used SPS models of an aging stellar population following a starburst to assign burst ages along the sequence. The red line and numbers are for representational purposes only. The circles represent our sample of 93 galaxies observed with Magellan/FIRE. Their colors correspond to their optical line ratio classification in the SDSS database (see section \ref{sec:sample:selection} for details). The left panels show examples of four optical spectra from our sample, where the SDSS identifiers are the plate-MJD-fiberID of each source. The spectra are ordered according to their location within the starburst to post-starburst sequence, showing continuous variation in their 4000~$\mathrm{\AA}$ break and Balmer absorption strengths as expected.}
\label{f:pca_selection_main_figure}
\end{figure*}

\section{The Starburst to Post-Starburst Sequence Sample}\label{sec:sample}

\subsection{Background}\label{sec:sample:background}

The optical spectrum of a galaxy encodes various properties of its underlying stellar population, such as the total stellar mass, star formation rate, stellar metallicity, initial mass function, and, of particular importance for this work, the star formation history (SFH; see review by \citealt{conroy13} and references therein). The spectrum around the 4000~$\mathrm{\AA}$ break has been used extensively to trace the mean age of the stellar population and the fraction of stars formed in the last $\sim$1 Gyr (e.g., \citealt{kauff03b, cid_fernandes05, gallazzi05}). In particular, the height of the 4000~$\mathrm{\AA}$ break and the strength of the Balmer absorption lines have been used to identify galaxies undergoing sharp transitions in their star formation rate, from early starbursts to quiescent post-starburst galaxies (e.g., \citealt{wild07, wild10}). 

The right panel of Figure \ref{f:pca_selection_main_figure} demonstrates the use of the 4000~$\mathrm{\AA}$ break and the Balmer absorption strengths in defining the starburst to post-starburst sequence. We show the distribution of $\sim$650\,000 local ($0.02 < z < 0.2$) galaxies with optical spectra from SDSS DR7 (\citealt{abazajian09}) in a two-dimensional plane that traces their 4000~$\mathrm{\AA}$ breaks and Balmer absorption strengths. This two-dimensional space is based on two spectral indices defined using a principal component analysis ({\sc pca}) of the 3175--4150 $\mathrm{\AA}$ spectral region of galaxy spectra (\citealt{wild07, wild10}). Galaxies with little to no star formation, which form the so-called `red sequence', are predominantly found on the right, showing strong 4000~$\mathrm{\AA}$ strengths corresponding to old stellar populations. Star-forming galaxies have weaker 4000~$\mathrm{\AA}$ break strengths, and are therefore located in the center/left of the diagram, with PCA1 indices around -4. 

\begin{figure*}
	\centering
\includegraphics[width=0.7\textwidth]{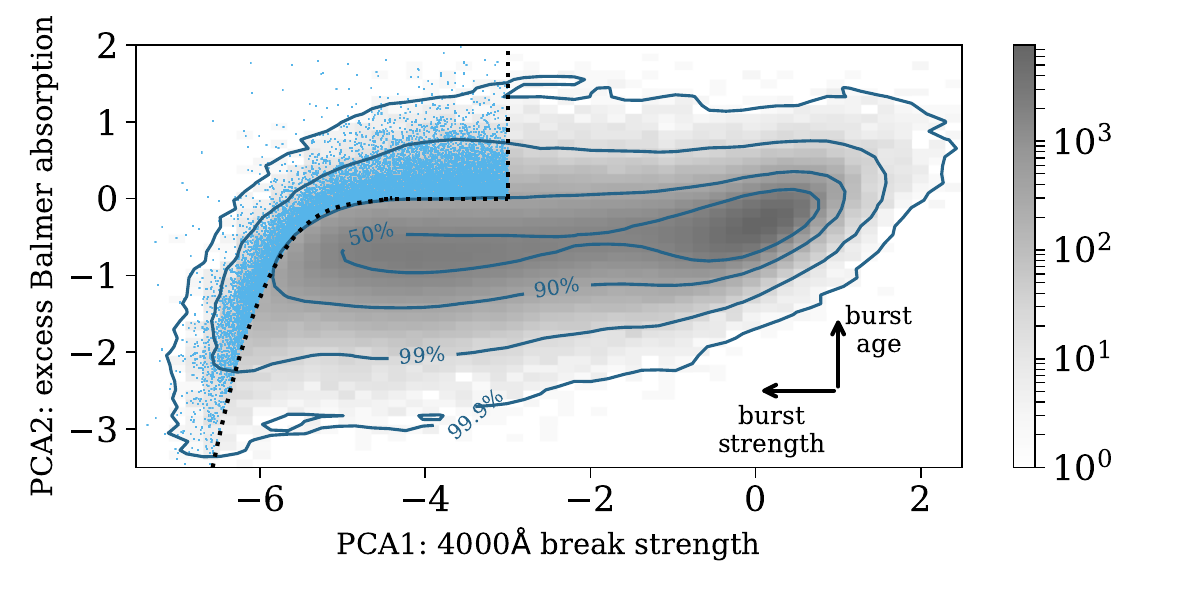}
\caption{\textbf{Selection of the parent sample based on the 4000~$\mathrm{\AA}$ break and Balmer absorption strengths.} The gray-scale shading shows the distribution of $0.02 \leq z < 0.17$ SDSS galaxies in the two-dimensional {\sc pca} plane defined by the 4000~$\mathrm{\AA}$ break and Balmer absorption strengths. Blue contours indicate various percentiles of the distribution. The evolution of a galaxy from starburst to post-starburst is expected to occupy the left-hand tail of the distribution (\citealt{wild07, wild09}). Galaxies above PCA2$=0$ are classified as Balmer-strong, with derived post-burst ages of $> 600$ Myr (\citealt{pawlik18}). We therefore include all galaxies with $\mathrm{PCA1} \leq -3$ and $\mathrm{PCA2} \geq 0$. For galaxies below PCA2$=0$, we select those lying outside the 90th percentile of the 2D distribution, following the criterion  $y = 0.48 x^{3} + 6.77 x^{2} + 31.86x +50.14$.  Light-blue points denote the galaxies that satisfy these selection criteria.
}
\label{f:parent_sample_selection}
\end{figure*}

\citet{wild07} identified the sequence from starburst to post-starburst as the left-most tail of the two-dimensional distribution, indicated in the figure with a red line, where time since the starburst increases from bottom to top, and the burst strength increases from right to left. Starburst galaxies are dominated by the ultraviolet-blue continua of O/B type stars, which have weak 4000~$\mathrm{\AA}$ strengths and weak Balmer absorption features, and are thus located in the bottom left. They differ from typical star-forming galaxies in the burst strength, making their spectra more dominated by the very young stars formed in the burst, moving them to the left tail of the distribution. Assuming that the star formation rate decreases rapidly after the peak of the burst, as the burst ages, the optical spectrum becomes dominated by the intermediate-age stellar population (A/F type stars), showing increased Balmer absorption strengths for a given 4000~$\mathrm{\AA}$ break. Using SPS models of an aging stellar population following a starburst, \citet{wild07} suggested that galaxies move along the red track as the starburst ages. The starburst ages indicated along the red track were assigned by \citet{wild10} assuming an exponentially-decaying SFH. These are shown for representation purposes only, and we derive our own starburst ages using {\sc prospector} fits of the optical-to-far infrared spectral energy distributions (SEDs) of the sources in our sample.

We use the two-dimensional plane defined by the 4000~$\mathrm{\AA}$ break and Balmer absorption strengths as the basis of our sample selection.

\begin{figure*}
	\centering
\includegraphics[width=1\textwidth]{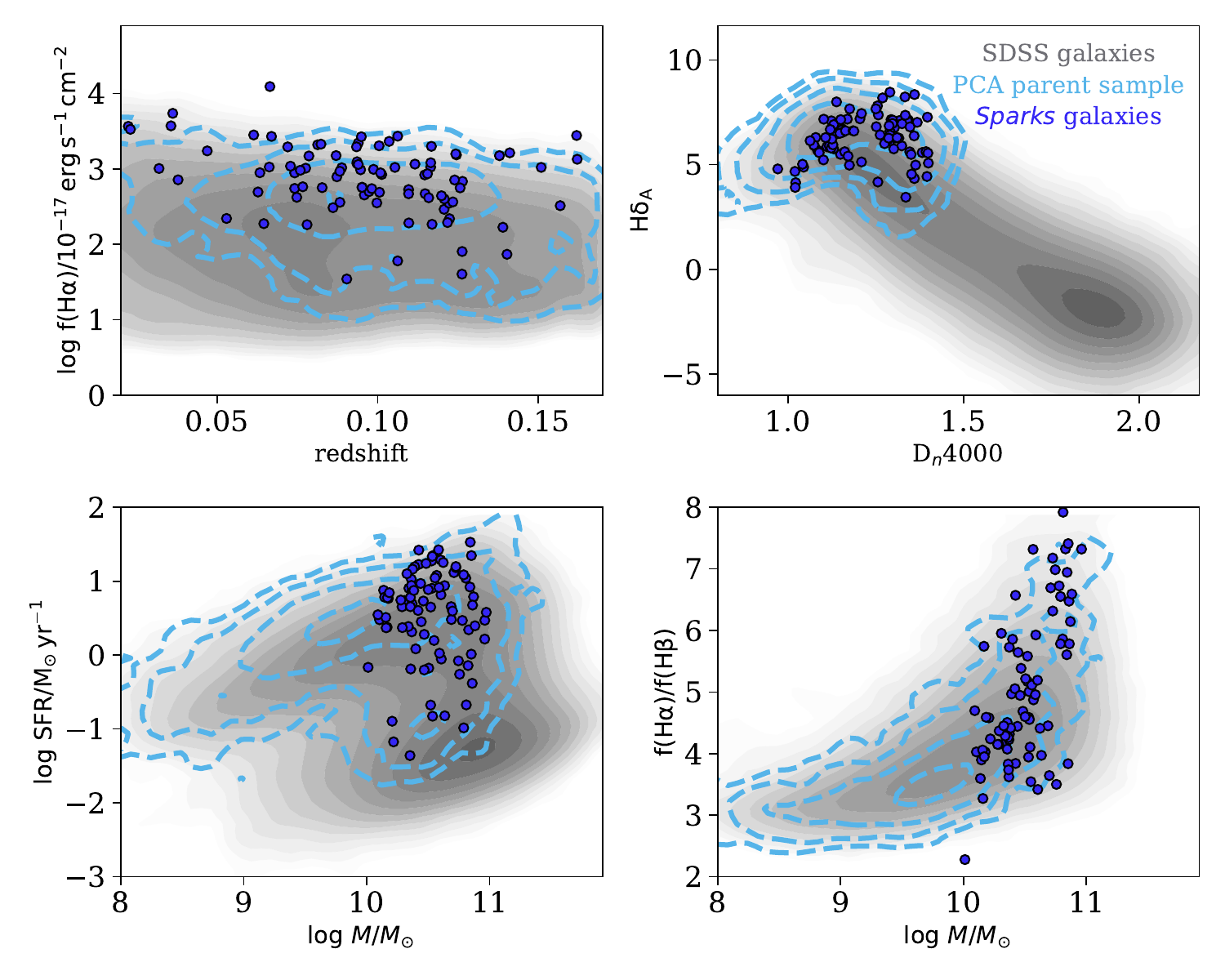}
\caption{\textbf{Comparison of general galaxy properties for SDSS, PCA-selected, and \textit{Sparks} galaxies}. The panels compare the \halpha flux versus redshift (top left); the $\mathrm{H\delta}_{\mathrm{A}}$ versus $D_{n}4000$ indices (top right); SFR versus stellar mass (bottom left), and \halpha/\hbeta flux ratio versus stellar (bottom right). The gray contours represent all SDSS galaxies within the redshift range $0.02-0.17$, and the light-blue contours represent the subset that passes the PCA selection criteria described in Section~\ref{sec:sample:selection}. Dark blue points represent the final \textit{Sparks} galaxy sample.}
\label{f:sample_properties_vs_SDSS_and_PCA}
\end{figure*}

\subsection{Sample selection and properties}\label{sec:sample:selection}

The sample we start from includes all the galaxies with optical spectra from SDSS DR7 \citep{abazajian09}, with available {\sc pca} amplitudes from \citet{wild07}\footnote{see \url{http://star-www.st-andrews.ac.uk/~vw8/downloads/DR7PCA.html}}. This includes the main galaxy sample and the luminous red galaxies sample. We cross-match it with the MPA-JHU catalog which provides SFR, stellar mass, and emission line flux measurements for SDSS galaxies (\citealt{kauff03b, b04, t04}). These properties are only used for sample selection, and in Paper~II, we fit SPS models to multi-wavelength photometry using multiple codes to derive final galaxy properties. 

Our goal is to select extreme systems that are located in the tail of the two-dimensional distribution defined by the 4000~$\mathrm{\AA}$ break and Balmer absorption strengths within the redshift range 0.02--0.17. The lower redshift limit is used to exclude galaxies that are too nearby and thus have large angular sizes that may cover the entire FIRE slit (size of 1\arcsec$\times$6\arcsec), preventing us from dithering between two on-source positions. The upper redshift limit is used to ensure that the near-infrared ro-vibrational H$_{2}$ line at $\lambda = 2.1218\, \mathrm{\mu m}$ is not redshifted outside the wavelength range covered by FIRE. 

Figure~\ref{f:parent_sample_selection} shows the 2D distribution of SDSS galaxies with $0.02 \leq z < 0.17$ in the {\sc pca} plane defined by the 4000~$\mathrm{\AA}$ break and Balmer absorption strengths. Using SPS models, \citet{wild07} and \citet{wild09} identified the left-hand tail of the distribution as the evolutionary track of a starburst galaxy, with post-burst age increasing from bottom to top, and burst strength increasing from right to left. Galaxies above $\mathrm{PCA2} = 0$ are considered Balmer-strong and typically have derived post-starburst ages above 600 Myr (e.g., \citealt{pawlik18}). We therefore include all galaxies with $\mathrm{PCA1} \leq -3$ and $\mathrm{PCA2} \geq 0$, where the limit on $\mathrm{PCA1}$ was set to roughly match the distribution of the galaxies by \citet{rowlands15} in {\sc pca} plane. For galaxies below $\mathrm{PCA2} = 0$, we select galaxies from the left-hand tail via the criterion $y = 0.48 x^{3} + 6.77 x^{2} + 31.86x + 50.14$, where $y$ and $x$ correspond to PCA2 and PCA1, respectively. This criterion was chosen to select the most extreme 10\% of galaxies in the {\sc pca} plane around $-1.5 <\mathrm{PCA2} < 0$, where the majority of the star-forming and quiescent galaxies lie. The criterion also approximately matches the {\sc pca} selection of starbursts to post-starbursts sequence by \citet{rowlands15}.  Out of 730\,464 SDSS galaxies with {\sc pca} measurements in this redshift range, 13\,618 satisfy the {\sc pca} selection criterion, of which 9\,515 have a median continuum SNR of 8 or higher.

\floattable
\begin{deluxetable}{cccc rrr rr c C}
\tablecaption{Magellan/FIRE galaxy sample\label{tab:galaxy_properties}}
\tablecolumns{10}
\tablenum{1}
\tablewidth{0pt}
\tablehead{
\colhead{(1)} & \colhead{(2)} & \colhead{(3)} & \colhead{(4)} & \colhead{(5)} & \colhead{(6)} & \colhead{(7)} & \colhead{(8)} & \colhead{(9)} & \colhead{(10)} & \colhead{(11)} \\
\colhead{plate} & \colhead{MJD} & \colhead{fiber ID} & \colhead{SpecObjID} & \colhead{R.A. (J2000)} & \colhead{Decl. (J2000)} & \colhead{redshift} & \colhead{PCA1} & \colhead{PCA2} & \colhead{BPT class} & \colhead{$\log M_{*}/M_{\odot}$} \\
}
\startdata
848 & 52669 & 76 & 954784056501889024 & 12:58:21.16 & +04:53:08.35 & 0.036 & -6.44 & 2.39 & 3 & 10.37^{+0.10}_{-0.08} \\ 
986 & 52443 & 561 & 1110291555639388160 & 21:17:01.46 & +00:40:58.73 & 0.094 & -5.51 & 0.83 & 1 & 10.37^{+0.13}_{-0.08} \\ 
644 & 52173 & 445 & 725201897132156928 & 21:51:28.37 & -06:46:30.70 & 0.101 & -4.11 & -0.62 & 4 & 10.52^{+0.13}_{-0.10} \\ 
1709 & 53533 & 287 & 1924241890027268096 & 14:29:44.99 & +09:32:36.05 & 0.122 & -4.37 & -1.03 & 5 & 10.86^{+0.13}_{-0.10} \\ 
\enddata
\tablecomments{(1)-(4) SDSS unique identifiers of the source, where SpecObjID is given for DR8 where the MPA-JHU catalog is available. (5)-(6) Right ascension and declination. (7) Redshift. (8)-(9) The amplitudes for the first two {\sc pca} components. (10) Galaxy class from the MPA-JHU catalog, where 1=`HII', 3=`composite', 4=`AGN', and 5=`weak lines'. (11) Stellar mass from the MPA-JHU catalog. A table containing all 93 sources is available online.
}
\end{deluxetable}

In Figure~\ref{f:sample_properties_vs_SDSS_and_PCA}, we show how the PCA-based selection criteria map onto key derived galaxy properties, including \halpha\ flux, redshift, $\mathrm{H\delta_{A}}$ and D$_n$4000 indices \citep{balogh99}, SFR, stellar mass, and \halpha/\hbeta\ flux ratios. By construction, in the $\mathrm{H\delta_{A}}$–D$_n$4000 plane, the parent sample includes only systems with low D$n$4000 values and elevated $\mathrm{H\delta_{A}}$ indices relative to the main star-forming locus. In the SFR–stellar mass plane, the parent sample includes starbursts galaxies that lie above the star-forming main sequence across $10^{8}$–$10^{11}\,\mathrm{M{\odot}}$, and post-starburst galaxies extending $\sim 1-2$ dex below the main sequence with stellar masses of  $\sim 10^{10}$–$10^{11}\,\mathrm{M{\odot}}$.

Since one of our goals is to study the connection between the starburst and the AGN, we focus on massive systems where the AGN fraction is larger (e.g., \citealt{kauff03a, goto07, mullaney13}). We use the SDSS-derived stellar mass by the MPA-JHU catalog (\citealt{kauff03b}) and select systems with masses in the range $10^{10}$--$10^{11}\,M_{\odot}$. We further restrict the sample to sources that can be observed with the Magellan telescope at Las Campanas Observatory, i.e., with Declination below 20$^{\circ}$. These additional criteria result in 1\,894 galaxies that constitute our parent sample.

The MPA-JHU catalog provides classifications of the sources according to their optical line ratios using standard line diagnostic diagrams (\citealt{baldwin81, veilleux87, kewley01, kauff03a}). The classes are: (i) galaxies with line ratios consistent with ionization by stars, marked `HII' throughout the paper; (ii) galaxies with line ratios consistent with AGN photoionization, marked `AGN'; (iii) galaxies that show a combination of stellar and AGN ionization, marked as `composite'; and (iv) galaxies showing weak or undetected lines (galaxies that do not satisfy the requirement of S/N$\, > 3$ in all four lines: \halpha, \hbeta, \oiii, and \nii; \citealt{b04}), marked `weak lines'. Within our parent sample, 1\,406 have detectable emission lines and 237 have weak or undetected lines. When selecting sources for follow-up observations with Magellan/FIRE, we include all the different galaxy classes.

\begin{figure}
	\centering
\includegraphics[width=\columnwidth]{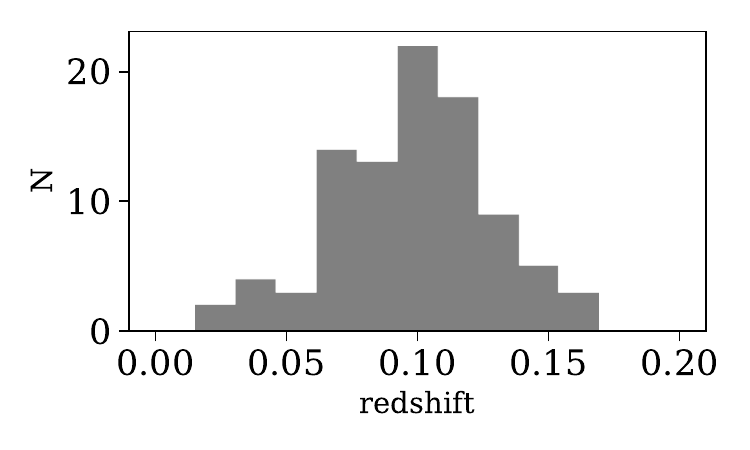}
\caption{\textbf{\emph{Sparks} galaxies redshift distribution.}}
\label{f:redshift_distribution}
\end{figure}

The primary goal of the program is the detection and analysis of various rest-frame near-infrared emission lines that trace multiphase gas. Therefore, for the galaxies in our parent sample with detected emission lines, we estimate the expected Pa$\alpha$ flux and prioritize systems with brighter lines within each galaxy class (`HII', `composite', and `AGN') and for different locations with respect to the star-forming main sequence. 
For galaxies with weak or undetected lines, we prioritize systems with brighter continua using their $K$-band magnitude. Since the prioritization is done in bins in the SFR-stellar mass plane, this does not introduce significant bias in global galaxy properties such as SFR, stellar mass, or dust reddening (\halpha/\hbeta), as can be seen in Figure~\ref{f:sample_properties_vs_SDSS_and_PCA}.

\begin{figure*}
	\centering
\includegraphics[width=1\textwidth]{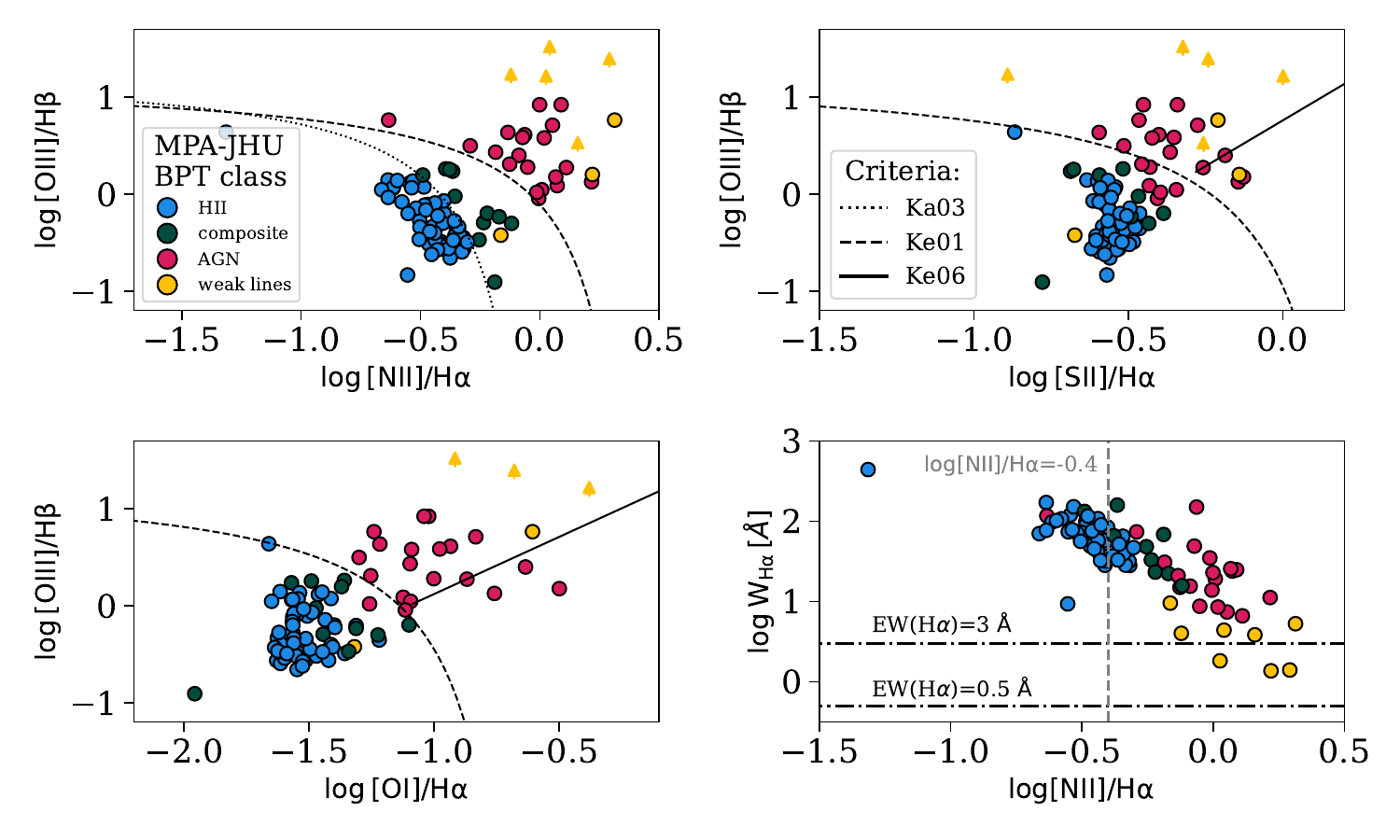}
\caption{\textbf{Location of the \emph{Sparks} galaxies in optical emission-line diagnostic diagrams} (\citealt{baldwin81, veilleux87}). The panels show the standard optical diagnostic diagrams, \oiiihbeta{} versus \niihalpha{}, \siihalpha{}, and \oihalpha{}, and EW(\halpha) versus \niihalpha (WHAN diagram; \citealt{cidfernandes11}), for the 93 galaxies in our sample, using emission-line measurements from the MPA-JHU catalog \citep{kauff03b, b04, t04}. Points are color-coded by the MPA-JHU spectral classification (see Section~\ref{sec:sample:selection}). Galaxies classified as `weak lines' typically have S/N$\,<3$ in H$\beta$, leading to uncertain \oiii/\hbeta ratios; we plot only galaxies with S/N$\,>3$ detections in all relevant lines, and indicate lower limits with arrows. The dividing curves commonly used to separate star-forming and AGN-dominated galaxies from \citet{kewley01} (Ke01) and \citet{kauff03a} (Ka03), as well as the LINER–Seyfert demarcation from \citet[Ke06]{kewley06}, are shown for reference.}
\label{f:optical_line_ratios_and_BPT}
\end{figure*}

From the parent sample, we observed 93 galaxies with Magellan/FIRE, selected at random based on observing availability. These galaxies constitute the \textit{Sparks} sample. We show their location in the two-dimensional {\sc pca} space in Figure \ref{f:pca_selection_main_figure}, along with four example optical spectra along the sequence. In Table \ref{tab:galaxy_properties} we list the general properties of these four galaxies. The full version of the table that includes all 93 Magellan/FIRE sources is available online. In Figure~\ref{f:redshift_distribution} we show their redshift distribution, and Figure~\ref{f:optical_line_ratios_and_BPT} presents their locations on several line diagnostic diagrams (BPT diagrams; \citealt{baldwin81, veilleux87, kewley01, kauff03a}). Out of the 93 galaxies in the sample, 53 are classified as purely star-forming (`HII'), 11 as composite, 21 as AGN-dominated, and 8 as weak-line systems. Most of the weak-line galaxies show detections in H$\alpha$, \nii, and \oiii, but only upper limits on H$\beta$. Depending on their (unknown) H$\beta$ flux, they may be classified as LINERs or Seyferts, but their line ratios are clearly inconsistent with being powered by star formation.

\section{Observational Program}\label{sec:fire}

This program uses the Folded port InfraRed Echellette (FIRE; \citealt{simcoe10}) spectrograph on the Magellan Baade Telescope. In its echelle mode, FIRE provides high-resolution ($\sim 50 \, \mathrm{km\,sec^{-1}}$) near-infrared spectra covering the wavelength range 0.8--2.5 \mic in a single setup. We use the FIRE echelle mode, with slit dimensions of $1\arcsec \times 6\arcsec$, to obtain rest-frame near-infrared spectra of the galaxies in our sample. In section \ref{sec:fire:obs} we describe the observations, and in section \ref{sec:fire:reduction} the data reduction. In section \ref{sec:fire:slit_losses} we estimate the extent of the line-emitting gas and derive correction factors to account for slit losses.

\subsection{Observations}\label{sec:fire:obs}

Observations were conducted over 20 observing nights from December 2022 to June 2024. In the afternoon of each observing run, we took dome flat-field calibration frames using the QL lamp. We observed between 2 and 7 galaxies per night, depending on the sources brightness and the weather conditions. Observations were mostly conducted in good weather conditions, with little to no cloud coverage and seeing between 0.5\arcsec\xspace and 1\arcsec. Since we observed local galaxies with angular extents that are larger than $\sim$1\arcsec, to minimize slit losses, we used the maximum slit width of 1\arcsec\xspace throughout the program. To minimize differential atmospheric refraction, the slit was oriented at the parallactic angle throughout the observations. For each source, we performed sequences of ABBA dithering, with exposure times ranging from 300 to 1200 seconds, depending on the source brightness, the phase of the moon and its proximity to the target, and the seeing. The separations between the A and B positions were 2.5\arcsec. We used the Sample Up The Ramp (SUTR) readout mode for the science targets. For each galaxy, we observed a telluric standard (A0V) star at an airmass comparable to that of the science target. For the telluric stars, we typically used AB dithering with exposure times ranging from 5 to 20 seconds with the Fowler 1 readout mode. 

\floattable
\begin{deluxetable}{ccc r ccc lcccc}
\tablecaption{Magellan/FIRE observations and line detections\label{tab:fire_obs}}
\tablecolumns{12}
\tablenum{3}
\tablewidth{0pt}
\tablehead{
\colhead{(1)} & \colhead{(2)} & \colhead{(3)} & \colhead{(4)} & \colhead{(5)} & \colhead{(6)} & \colhead{(7)} & \colhead{(8)} & \colhead{(9)} & \colhead{(10)} & \colhead{(11)} & \colhead{(12)} \\
\colhead{plate} & \colhead{MJD} & \colhead{fiber ID} & \colhead{Exposure time (sec)} & \colhead{\#H} & \colhead{\#Fe} & \colhead{\#H$_{2}$} & \colhead{Slit-loss correction?} & \colhead{FWHM(Pa$\alpha$) (\arcsec)} & \colhead{$1/f_{\mathrm{corr}}$} & \colhead{$(1/f_{\mathrm{corr}})_{\mathrm{lower}}$} & \colhead{$(1/f_{\mathrm{corr}})_{\mathrm{upper}}$}\\
}
\startdata
848 & 52669 & 76 & 5400 & 5 & 2 & 3 & 1 & 0.9 & 0.69 & 0.40 & 0.96 \\ 
986 & 52443 & 561 & 4800 & 3 & 0 & 0 & 0 & 1.3 & - & - & - \\ 
644 & 52173 & 445 & 6600 & 2 & 0 & 0 & 1 & 1.1 & 0.49 & 0.27 & 0.78 \\ 
1709 & 53533 & 287 & 8400 & 1 & 0 & 3 & 1 & 1.0 & 0.66 & 0.37 & 0.93 \\ 
\enddata
\tablecomments{(1)-(3) SDSS plate, MJD, and fiber ID. (4) Total on-source exposure time. (5) Number of detected hydrogen emission lines. (6) Number of detected iron lines. (7) Number of detected H$_{2}$ lines. A table containing all 93 sources is available online. Line detection statistics for the full sample are also presented in Figure \ref{f:line_detection_vs_MS_and_PCA}. (8) A flag indicating whether it is safe to use the Pa$\alpha$ slit correction factor to obtain the total Pa$\alpha$ flux. The flag is `0' in cases where a substantial part of the line profile is outside of the slit. (9) The FWHM size of the Pa$\alpha$ emitting gas, estimated from fitting Gaussians to the emission along the slit. The values are not corrected for seeing. (10) The inverse correction factor for slit losses assuming the source is circularly symmetric. (11) and (12) The inverse correction factor assuming the source is $\times 2$ and $\times 1/2$ more extended along the slit 1\arcsec\xspace dimension. 
}
\end{deluxetable}

The total exposure times per source are given in table \ref{tab:fire_obs}, and they range between 1200 seconds and 13\,200 seconds.

\begin{figure*}
	\centering
\includegraphics[width=1\textwidth]{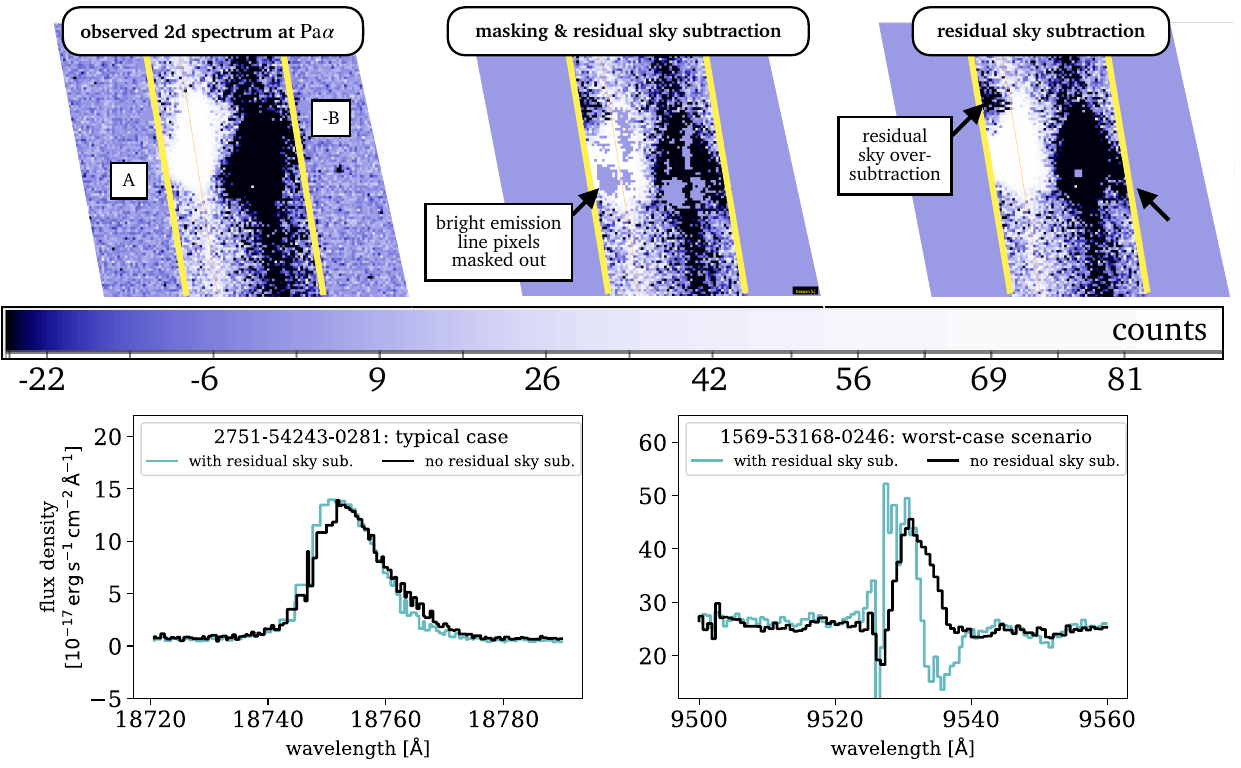}
\caption{\textbf{Impact of masking and residual sky subtraction on strong and extended emission lines.} The top row includes three panels showing the two-dimensional combined A-B-B+A spectra around the Pa$\alpha$ line of an extreme galaxy in its spatially-extended Pa-$\alpha$ emission. In most of our sample, the A and B spectra are more compact and do not cancel each other. The yellow lines represent the edges of the slit orders. The white (black) colors represent positive (negative) counts and correspond to the A (B) position. The left panel shows the observed spectrum. The middle panel shows the spectrum after masking (\texttt{use\_2dmodel\_mask=True}) and residual sky subtraction (\texttt{no\_local\_sky=False}) have been applied, which are part of the default \texttt{PypeIt} setting. The right panel shows the 2d spectrum when masking is turned off (\texttt{use\_2dmodel\_mask=False}) but residual sky subtraction is applied. The local sky model includes a contribution of the extended emission line, and sky subtraction results in over-subtraction of the emission line. For each object in the sample, we set \texttt{use\_2dmodel\_mask=False} and reduce the observations with and without \textit{residual} sky subtraction. The bottom row compares the resulting 1d spectra with and without residual sky subtraction for a typical galaxy (left panel; same galaxy as in the top row) and one of the worst cases in our sample (right panel; SDSS ID: 1569-53168-0246; highly extended galaxy). In both bases, masking is turned off.}
\label{f:reduction_fig}
\end{figure*}

\subsection{Data reduction}\label{sec:fire:reduction}

All data reduction steps are performed with version 1.13 of the \texttt{PypeIt} reduction pipeline \citep{prochaska20}. For each galaxy, we use the dome flats to trace the slit order edges and to correct for pixel-to-pixel detector variations. \texttt{PypeIt} uses the observed sky OH lines to obtain a wavelength solution, and thus we use the science frames as arc and tilt frames. Science exposures are grouped into groups of four, corresponding to the ABBA dithering, where separate calibration files are produced for each group. The two telluric standard exposures are grouped to the four science exposures taken closest in time to them. Within each group, science exposures are combined as A-B-B+A. Despite the difference imaging (i.e., the A-B), there are significant sky residuals in the combined frames. The pipeline then performs object finding, residual sky subtraction, and object extraction on the combined frames. 

Most of the galaxies are bright enough to show a detectable trace throughout the different orders, so that the standard object finding routines work well in our case. Many of the galaxies exhibit strong emission lines that can have different spatial profiles than that of the continuum. In extreme cases where the emission line profiles differ substantially from the continuum profile, some emission line pixels are flagged and masked out (see illustration in the middle panel of Figure \ref{f:reduction_fig}). To avoid this, we set \texttt{use\_2dmodel\_mask=False} in the object extraction stage for all the sources. 

\begin{figure*}
	\centering
\includegraphics[width=1\textwidth]{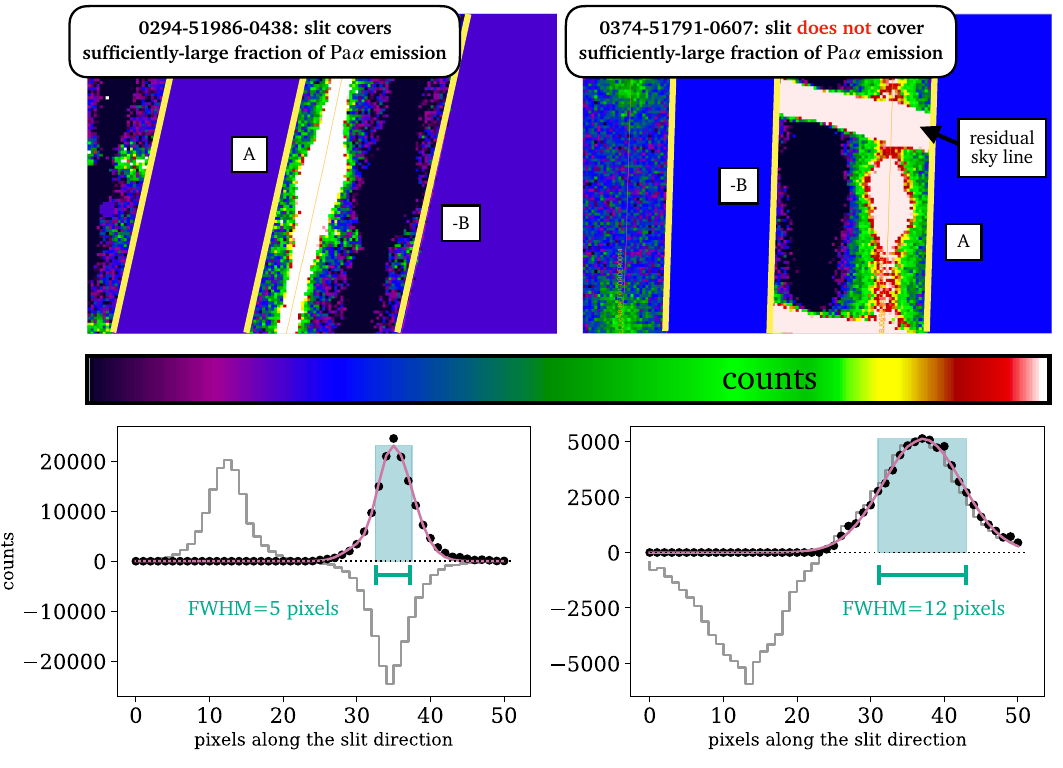}
\caption{\textbf{Using the Pa$\alpha$ emission along the slit direction to estimate the angular extent of the line-emitting region.} The top row shows the 2d reduced spectra by \texttt{PypeIt} centered around the Pa$\alpha$ emission line. The color-coding represents the counts, where white (black) colors corresponds to positive (negative) count values associated with the A (B) position. The yellow lines represent the slit order edges. The left panel shows a typical galaxy in the sample with a relatively compact line-emitting region (SDSS ID: 0294-51986-0438) and the right panel shows a worst-case scenario example of an extended line region (SDSS ID: 0374-51791-0607). The bottom panels show the counts of the Pa$\alpha$ along the slit direction (gray solid line). Splitting the negative and positive counts at half the slit width (pixel 25), we take the absolute value of the counts (indicated with black points), and fit the counts with a sum of 2 Gaussians (pink solid line), and estimate the FWHM of the line-emitting region (green boxes and text). In section \ref{sec:fire:slit_losses} we use these extents to derive slit loss correction factors for systems with sufficiently compact line emitting regions.}
\label{f:slit_corrections}
\end{figure*}

For the subtraction of residual sky lines, \texttt{PypeIt} builds a local model of the sky using pixels that are close to the edges of the slit. For the galaxies in our sample that have spatially extended emission lines, the local sky model can include a contribution from the galaxy's emission lines, which is then subtracted from the object's model during the extraction stage. This is illustrated in Figure~\ref{f:reduction_fig} which shows an extreme case of spatially-extended Pa-$\alpha$ emission. We note that in most of our sources, the Pa-$\alpha$ emission is much more compact and that A and B dithers are not canceling each other. We examined different solutions to this over-subtraction, including changing the extraction method from optimal extraction to boxcar extraction, and found that the issue persists. In such cases, the only solution is to turn off the local sky model altogether, where no additional sky subtraction is performed (setting \texttt{no\_local\_sky=True} in the sky subtraction routine). Although it results in spectra with a significant contamination from residual sky lines, the emission lines are extracted correctly. In most cases, the residual sky lines do not coincide in wavelength with the emission lines from the galaxy. In the rare cases that they do, they are substantially narrower than the emission lines from the galaxy, and can therefore be masked out manually.

For each of the galaxies in the sample, we run the reduction pipeline twice, once with sky subtraction, and once with the sky subtraction turned off. At the end of this stage, we have one-dimensional (1d) spectra extracted from each of the combined frames. We use the 1d spectrum of the telluric standard to perform flux calibration, where \texttt{PypeIt} first constructs a sensitivity function and then applies it to all the extracted 1d spectra of the galaxy and star. We then coadd all the 1d spectra and apply a telluric correction using the observed telluric standard. 

\subsection{Extent of the line-emitting gas and slit losses}\label{sec:fire:slit_losses}

The galaxies in our sample are local and typically have optical sizes larger than 1\arcsec. In this section we use the reduced 2d spectra and analyze the Pa$\alpha$ emission along the slit direction. By fitting the Pa$\alpha$ emission profile along the 6\arcsec-long slit, we estimate the angular extent of the line-emitting gas. For the subset of the galaxies where the FIRE 1\arcsec-wide slit covers most of the line-emitting region, we estimate the Pa$\alpha$ flux outside the slit and derive slit correction factors. The process is illustrated in Figure \ref{f:slit_corrections}, where we show an example of a typical galaxy in the sample with a relatively compact line-emitting region, and a worst-case scenario example of a galaxy with an extended line region that is not covered sufficiently well by FIRE. 

Out of the 93 galaxies in our sample, 83 are detected in Pa$\alpha$ with S/N$\,>5$. For these sources, we use the reduced 2d spectra products by \texttt{PypeIt}, which include the identified slit order edges, the 2d wavelength solution, and a mask that identifies cosmic rays and other artefacts in the 2d detector images. We extract the counts in the wavelength range $\lambda = [18750 - 5, 18750 + 5]\, \mathrm{\AA}$ using the wavelength solution, converted to air wavelength and shifted to rest-frame using the SDSS redshift. Our manual inspection of the 2d spectra shows that this wavelength range includes most of the Pa$\alpha$ emission in the majority of the sources. For six of the galaxies, the Pa$\alpha$ emission is redshifted with respect to 18750 \AA\xspace assuming the SDSS redshift. For these sources, we manually change the wavelength range to include most of the Pa$\alpha$ emission. There are about 50 pixels along the slit direction, and we sum the counts of all the pixels that are not masked-out within this wavelength range. This results in a one-dimensional vector that represents the counts as a function of position along the slit, where the counts are positive around the A position and negative around the B position. These are shown with gray lines in Figure~\ref{f:slit_corrections}.

Next, we model the emission along the slit by fitting a sum of two Gaussians to the positive/negative counts separately. The splitting between the positive and negative counts is done at the half of the slit (pixel 25). Then, for each separately, we take the absolute value of the counts (shown with black points in Figure~\ref{f:slit_corrections}) and fit a sum of two Gaussians. We estimate the full width at half maximum (FWHM) of the best-fitting profile, using the peak of the summed Gaussians as the reference point. The FWHM is measured in pixels, and we convert it to \arcsec\xspace via $\mathrm{FWHM}(\arcsec)= \mathrm{FWHM(pix)} \times (6\arcsec/N_{\mathrm{pix}})$, where $N_{\mathrm{pix}}$ is the number of pixels along the slit direction. Using the derived FWHM, we define a criterion of whether the FIRE slit covers a sufficiently-large fraction of the Pa$\alpha$ emission. We define $\mathrm{\sigma(\mathrm{pix})=\mathrm{FWHM(pix)}/2.355}$. For galaxies with $6 \times \sigma(\mathrm{pix}) < N_{\mathrm{pix}}/2$, we consider the FIRE slit to cover a sufficiently-large fraction of the Pa$\alpha$ flux to derive a slit correction factor. For these galaxies, more than 95\% of the Pa$\alpha$ flux along the slit is within 3\arcsec. Galaxies with $6 \times \sigma(\mathrm{pix}) > N_{\mathrm{pix}}/2$ are considered to be too extended for a derivation of an accurate slit correction factor. Out of the 83 galaxies, 63 pass our criterion ($6 \times \sigma(\mathrm{pix}) < N_{\mathrm{pix}}/2$) and 20 do not.

\begin{figure*}
	\centering
\includegraphics[width=1\textwidth]{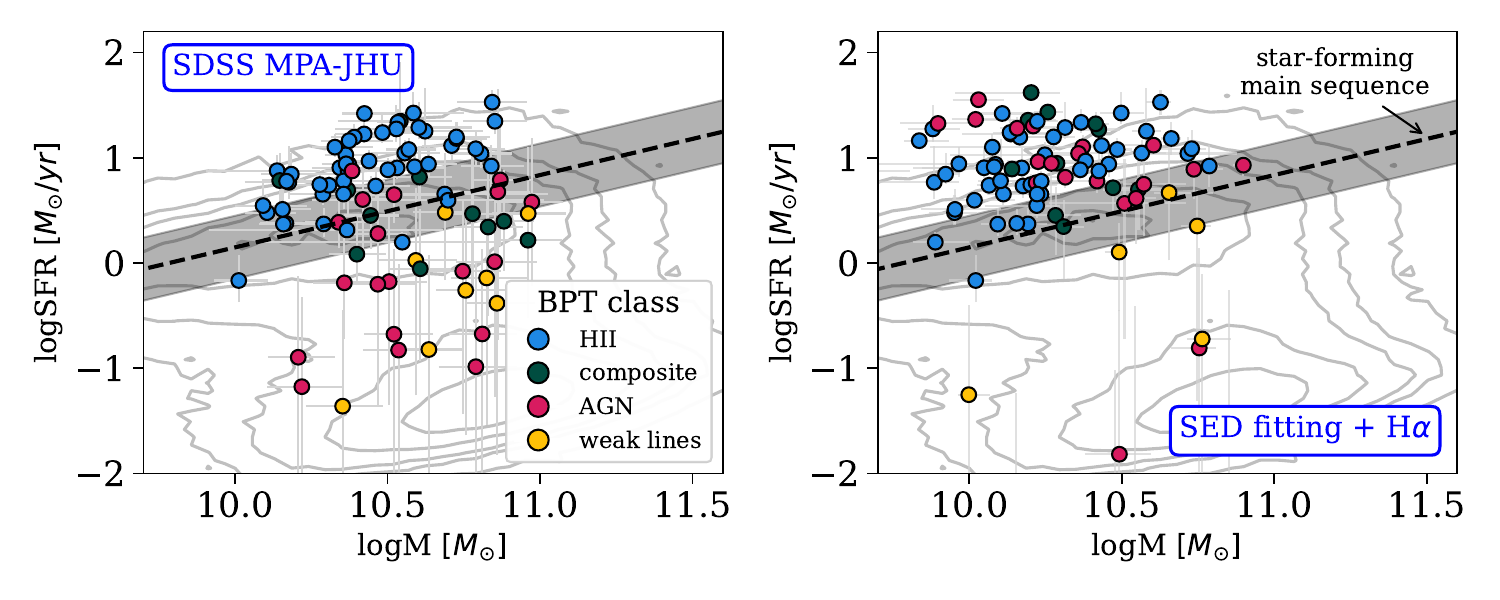}
\caption{\textbf{Star formation rate versus stellar mass for the \emph{Sparks} galaxies.} The left panel shows galaxy properties from the MPA-JHU catalog. The right panel shows our adopted properties using a combination of H$\alpha$ flux and SED fitting (see section \ref{sec:results}): SFR is derived using either the \halpha flux or the best-fitting \texttt{Prospector} fits to the multi-wavelength SED, and stellar mass is based on \texttt{Prospector} fits. Points are color-coded by the MPA-JHU spectral classification. For comparison, the gray contours represent local \emph{SDSS} galaxies, and the black dashed line represents the star forming main sequence at z=0.1 by \citet{whitaker12} with shaded regions showing $\pm 0.3$ dex.}
\label{f:SFR_vs_stellar_mass}
\end{figure*}

For the 63 galaxies that pass our criterion, we estimate the fraction of Pa$\alpha$ emission that is not covered by the 1\arcsec-wide slit and derive a slit correction factor. We use the best-fitting profile and assume that it represents a two-dimensional circularly symmetric double Gaussian profile. We integrate the entire profile to obtain $I\mathrm{(total\,profile)}$, and integrate over an area of $1 \arcsec \times 6 \arcsec$ to obtain $I\mathrm{(slit)}$. We define the inverse of the correction factor to be $I\mathrm{(slit)}/I\mathrm{(total \, profile)}$. For the galaxies that pass our criterion, the minimum inverse correction factor for a circularly symmetric profile is 0.36, the maximum is 0.99, and the average is 0.66. Since the line emitting regions can be asymmetric, we derive two additional correction factors, assuming that the regions are $\times 2$ and $\times 1/2$ more extended in the 1\arcsec\xspace slit direction. We do not consider the case of more asymmetric profiles since manual inspection of the SDSS images of our sources shows that the systems are generally face-on, showing quite symmetric profiles in their optical broad-band images. These corrections differ by a factor of $< 2$ from the correction derived assuming circularly symmetric profiles. We therefore suggest to adopt an uncertainty of $\sim$0.2--0.3 dex on measurements that require the total integrated near-infrared line emission flux (e.g., the Pa$\alpha$ derived SFR). In Table \ref{tab:fire_obs} we indicate for each galaxy whether it passes our criterion, as well as list the derived FWHM of the Pa$\alpha$ and the inverse of the correction factors. 

\section{Refining the location along the starburst to post-starburst sequence with multi-wavelength information}\label{sec:results}

The \textit{Sparks} galaxies were initially selected based on physical properties derived from optical spectroscopy and photometry (see Section~\ref{sec:sample}). Specifically, (i) their positions along the starburst to post-starburst sequence were defined using a {\sc pca} decomposition of optical spectra around the 4000~$\mathrm{\AA}$ break, with each galaxy’s evolutionary stage inferred from its excess Balmer absorption at a given 4000~$\mathrm{\AA}$ break; and (ii) their locations in the SFR–$M_{*}$ plane relative to the star-forming main sequence were determined using the SFRs and stellar masses from the MPA–JHU catalog \citep{kauff03b, b04, t04}, based solely on optical information.
In this section, we refine the placement of the \textit{Sparks} galaxies within the sequence using new measurements of SFRs, stellar masses, dust reddening, and star formation histories derived from SPS fits to the optical spectra and multi-wavelength photometry (Paper~II). The inclusion of the multi-wavelength photometry provides improved constraints on stellar masses (via rest-frame near-infrared fluxes) and on the ongoing SFR and dust obscuration of young massive stars (via ultraviolet and mid- to far-infrared emission).

Since our analysis derives new SFRs and stellar masses, we first summarize how these quantities were estimated in the MPA–JHU catalog \citep{kauff03b, b04, t04}. Stellar masses in the catalog were derived from SPS fits to the \textit{SDSS} $ugriz$ photometry only. Multiple studies have demonstrated that stellar mass estimates based solely on optical photometry are subject to systematic uncertainties and offsets compared with fits including near‐infrared data (e.g., \citealt{salim07, conroy13, leja19, lower20, tacchella22}; and Paper~II). SFRs were estimated using either the H$\alpha$ line flux or the D${_n}$4000 \AA\, index, depending on galaxy type \citep{b04}. For galaxies with line ratios consistent with pure star formation, the SFR was calculated from the dust-corrected H$\alpha$ flux, further corrected for aperture losses from the 3\arcsec\ \textit{SDSS} fiber. For galaxies with line ratios inconsistent with pure star formation (composites, AGN, and weak-line systems), the SFR was estimated from the D${_n}$4000 \AA\, index, calibrated using the pure star-forming galaxies using their \halpha flux. The D$_{n}$4000 \AA\ index-based SFR is highly uncertain, with a scatter of $\sim$0.4 dex relative to H$\alpha$-based SFRs in the general star-forming galaxies population. Since the \textit{Sparks} galaxies are extreme in their D$_{n}$4000 $\mathrm{\AA}$ and Balmer absorption properties, the 0.4 dex scatter may not represent them well. The left panel of figure \ref{f:SFR_vs_stellar_mass} shows the SFR versus stellar mass of the \textit{Sparks} galaxies using the properties listed in the MPA-JHU catalog. Our goal is to improve these estimates by including multi-wavelength photometry.

\begin{figure*}
	\centering
\includegraphics[width=1\textwidth]{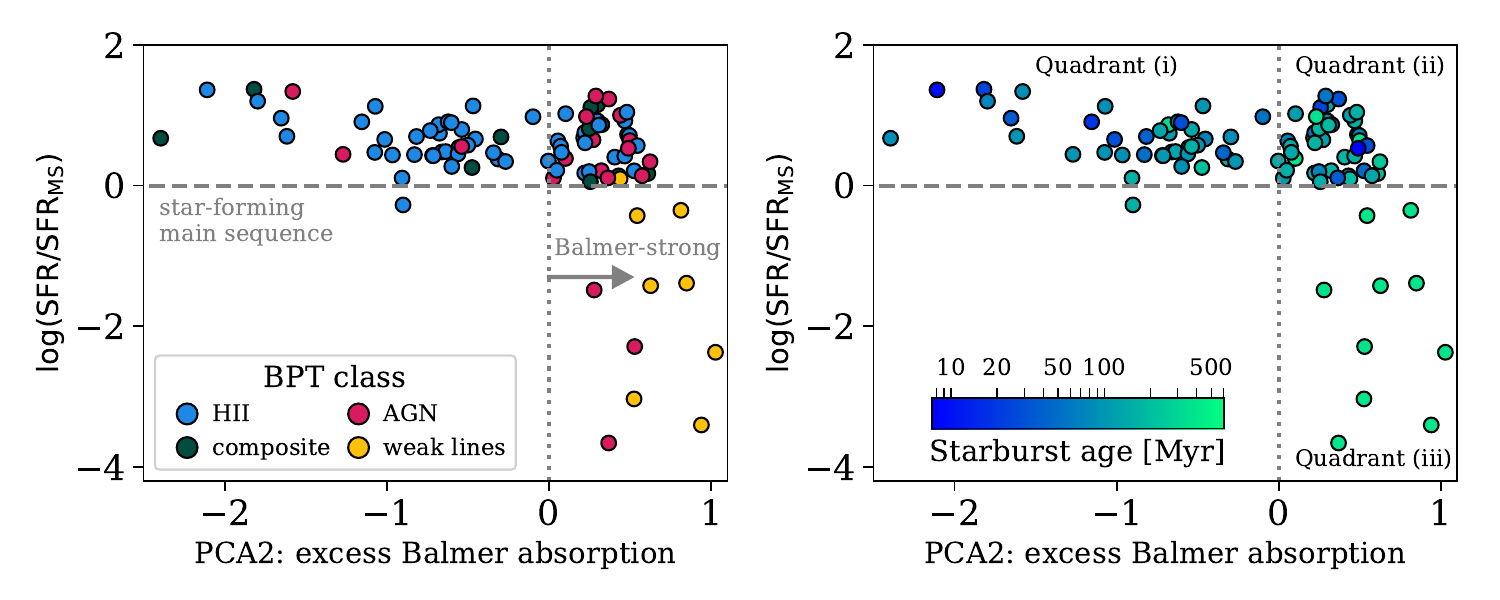}
\caption{\textbf{Offset from the star-forming main sequence versus PCA2 for the \emph{Sparks} galaxies.} The y-axis shows the offset from the star-forming main sequence, $\log (\mathrm{SFR}/\mathrm{SFR}_{\mathrm{MS}})$, where $\mathrm{SFR}_{\mathrm{MS}}(M_*, z)$ is taken from \citet{whitaker12} and the adopted SFRs are described in Section~\ref{sec:results}. The x-axis shows the PCA2 coefficient, which measures the excess Balmer absorption relative to the D$_n$4000 $\mathrm{\AA}$ index and traces the time since the last starburst \citep{wild07}. The horizontal dashed line marks the star-forming main sequence, and the vertical dotted line indicates the PCA2 threshold above which galaxies are classified as Balmer-strong \citep[e.g.,][]{pawlik18}, corresponding to $\gtrsim 600$ Myr since the burst. In the left panel, galaxies are color-coded by MPA-JHU spectral classification (Section~\ref{sec:sample:selection}); in the right panel, by the starburst age derived from \texttt{Prospector} fits to the multi-wavelength SEDs.}
\label{f:Delta_MS_vs_PCA2}
\end{figure*}

We compiled far-ultraviolet to far-infrared photometry for the \textit{Sparks} galaxies and applied several SPS modeling frameworks to derive their physical properties. In particular, the multi-wavelength spectral energy distributions (SEDs) were fitted with \texttt{MAGPHYS} \citep{daCunha08} and \texttt{Prospector} \citep{leja17}. We additionally fitted the optical spectra with \texttt{Prospector}, adopting the same model ingredients and priors as in the multi-wavelength SED fits to ensure consistency. These procedures are described in detail in Paper~II, where we compare results across modeling codes, datasets, and assumptions; identify the set of physical parameters we adopt as the final properties for the \textit{Sparks} galaxies; and assess their uncertainties. Here, we summarize only the adopted physical properties relevant to the survey.


Our adopted SFRs and stellar masses are described below. The main difference with respect to the MPA-JHU properties is that the new estimates are based on fitting far-ultraviolet to far-infrared photometry, rather than $ugriz$ photometry or the D$_{n}$4000 $\mathrm{\AA}$ index. For galaxies with optical line ratios consistent with pure star-formation ionization, the H$\alpha$ flux is powered by young, massive stars and provides the most reliable tracer of the instantaneous SFR. In galaxies with line ratios inconsistent with pure star formation, however, the H$\alpha$ emission may be powered by other processes (e.g., AGN or shocks) and thus cannot be used as an SFR indicator. Using \texttt{Prospector} SED fits to the far-ultraviolet to far-infrared photometry, we derived average SFRs over the past 10 Myr. For purely star-forming galaxies, the SED-based SFRs show reasonable agreement with the \halpha-based SFRs (offset $\sim$$-0.15$ dex, scatter $0.14$ dex; Paper~II). This is in contrast with the SFRs derived from fitting the optical continuum emission (which includes the D$_{n}$4000 $\mathrm{\AA}$ index), which substantially under-predicts the ongoing SFR based on \halpha (offset of $-0.76$ dex, scatter of $0.42$ dex; Paper~II). We attribute the better agreement with the SED-based SFRs to the inclusion of mid- and far-infrared photometry, which constrains the contribution of the youngest stellar populations. 

Accordingly, we adopt H$\alpha$-based SFRs for the purely star-forming galaxies, and the 10 Myr-averaged SFRs from \texttt{Prospector} SED fits for composites, AGN hosts, and weak-line systems. We do not correct for the -0.15 dex offset between SFR(SED; 10 Myr) and SFR(H$\alpha$) we observed for the star-forming galaxies. Thus, our adopted SFRs are identical to those by the MPA-JHU catalog for star-forming galaxies, while for composites and AGN we use SED-based SFRs instead of D$_{n}$4000 $\mathrm{\AA}$ index-based. For consistency, we adopt stellar masses derived from the same \texttt{Prospector} model for all \textit{Sparks} galaxies. These show $-0.2$ dex offsets and $0.15$ dex scatter compared to the MPA-JHU stellar masses.

The Pa-$\alpha$ line covered by our FIRE observations provides an additional constraint on the SFR that is less affected by dust extinction. As with H$\alpha$, for Pa-$\alpha$ to reliably trace ongoing star formation, near-infrared line diagnostic diagrams must indicate that the emission is powered by young, massive stars rather than by shocks or an AGN. For the \textit{Sparks} galaxies, we identify cases where the optical and near-infrared line ratios yield conflicting classifications of the primary ionizing source. Consequently, we do not include Pa-$\alpha$-based SFRs in the present analysis; the origin of the discrepancy between optical and near-infrared diagnostics will be addressed in a forthcoming paper.

In the right panel of Figure \ref{f:SFR_vs_stellar_mass} we show our final adopted SFRs versus stellar masses for the \textit{Sparks} galaxies. The largest differences from the initial SDSS-based estimates occur for composites and AGN hosts. In the MPA-JHU catalog, most of these galaxies fall below the star-forming main sequence, whereas our SED-fitting–based estimates place them above the sequence. If the position relative to the star-forming main sequence reflects evolutionary stage, our revised estimates suggest that composites and AGN hosts are at an earlier stage of evolution than implied by the D$_{n}$4000 \AA-based estimates—reclassifying them from being on their way to quenching to being actively forming stars. 

\begin{figure*}
	\centering
\includegraphics[width=1\textwidth]{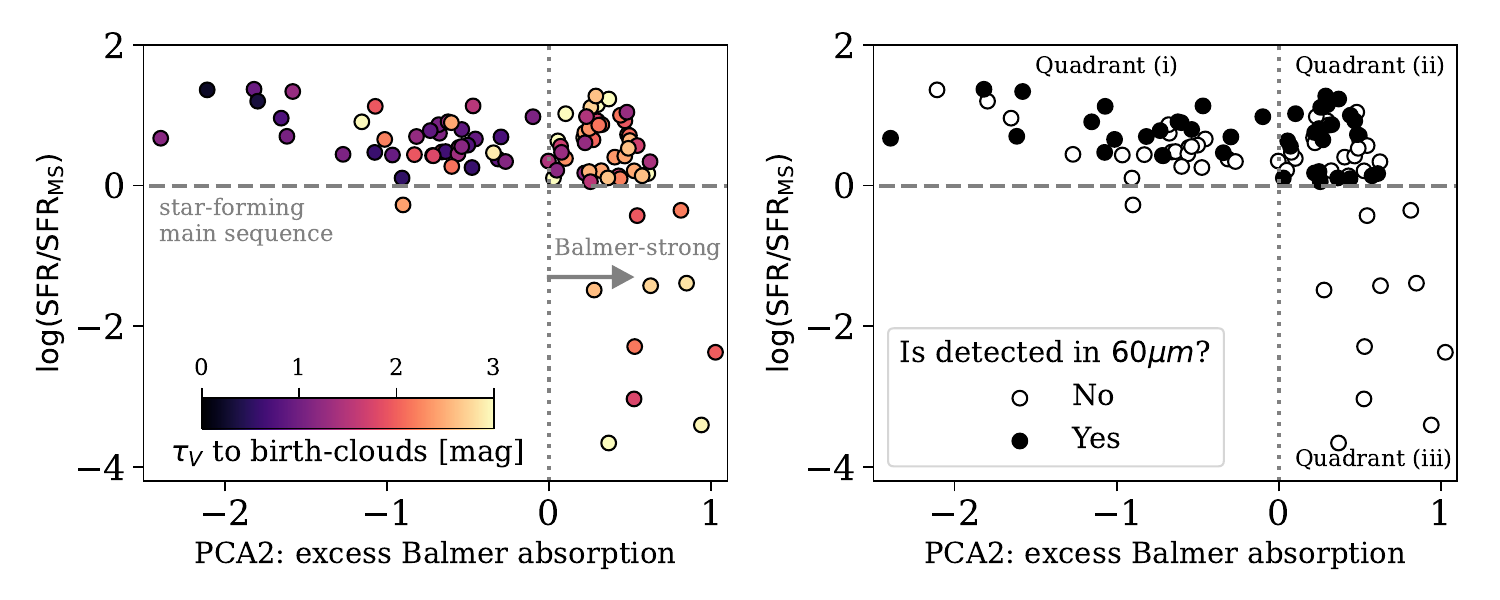}
\caption{\textbf{Dust reddening and far-infrared detection across the $\Delta(\mathrm{MS})$--PCA2 plane.} Axes are the same as in figure~\ref{f:Delta_MS_vs_PCA2}. The horizontal dashed line marks the star-forming main sequence, and the vertical dotted line indicates the PCA2 threshold above which galaxies are classified as Balmer-strong. In the left panel, points are color-coded by the dust optical depth towards the youngest stellar populations, $\tau_{V}$, from the free obscuration model. In the right panel, points are color-coded by whether or not far-infrared emission is detected by IRAS at $60\,\mathrm{\mu m}$.}
\label{f:Delta_MS_vs_PCA2_dust_and_IR}
\end{figure*}

A more direct probe of the SFH over the past several hundred Myr can, in principle, be obtained from SPS fitting of the optical spectroscopy tracing stellar continuum. Many of the \textit{Sparks} galaxies show strong Balmer absorption lines, a hallmark of intermediate-age stellar populations, so fitting their optical continuum can potentially provide tight constraints on the timing and strength of a recent or ongoing burst. In Paper~II, we fitted the optical continuum using \texttt{Prospector} with similar model ingredients and priors to those adopted for the \texttt{Prospector} SED fitting. The best fits result in SFR(10 Myr) values offset by $-0.76$ dex (scatter $0.42$ dex) relative to H$\alpha$-based SFRs in star-forming galaxies, and they systematically underpredict the 60 \mic flux by $0.57$ dex (scatter $0.15$ dex). Lacking constraints in ultraviolet and far-infrared, the optical continuum fits tends to prefer intermediate-age stellar populations to fit the strong Balmer absorption, while underestimating the ongoing star formation. For the purposes of this paper, we therefore adopt SFH parameters from the \texttt{Prospector} multi-wavelength SED fits, which we consider to trace the SFH more robustly than the optical continuum fits. In particular, we estimate the starburst age, $t_{\mathrm{SB}}$, defined as the lookback time when 90\% of the stellar mass formed within the last 1~Gyr.


In Figure~\ref{f:Delta_MS_vs_PCA2} we show the offset from the star-forming main sequence versus the PCA2 coefficient for the \textit{Sparks} galaxies. PCA2 quantifies the excess Balmer absorption relative to the D$_n$4000 \AA\ index and has been proposed as a tracer of time since the last starburst \citep{wild07}. Galaxies with PCA2 $>0$ are considered Balmer-strong, corresponding to post-burst ages of $\gtrsim 600$ Myr \citep{wild07, pawlik18}. The \textit{Sparks} galaxies form an L-shaped distribution in this space, which we divide into three quadrants: (i)~$\Delta(\mathrm{MS}) > 0$ and PCA2 $<0$, actively star-forming galaxies without significant Balmer absorption; (ii)~$\Delta(\mathrm{MS}) > 0$ and PCA2 $>0$, galaxies above the main sequence (by up to 1 dex) but still showing Balmer absorption excess; and, (iii)~$\Delta(\mathrm{MS}) < 0$ and PCA2 $>0$, galaxies below the main sequence with strong Balmer absorption. As expected, many star-forming galaxies fall in quadrant (i), while most weak-line galaxies lie in (iii). Strikingly, composites and AGN hosts populate quadrant (ii), where SED-based SFRs imply substantial ongoing star formation\footnote{Since we did not correct the SFRs of composites and AGN hosts for the $-0.15$ dex offset between SFR(10 Myr; SED) and SFR(H$\alpha$), their ongoing star formation may in fact be even higher.}, yet the stellar continuum indicates the presence of a substantial post-burst population.

In the right panel of Figure~\ref{f:Delta_MS_vs_PCA2}, we color-code the galaxies by their starburst ages. Overall, PCA2 increases with $t_{\mathrm{SB}}$, but with several notable exceptions, primarily in quadrant (ii). Some star-forming galaxies lie above the main sequence while also showing Balmer-strong continua, with starburst ages of $\sim 100-200$ Myr. In contrast, some composites and AGN hosts in quadrant (ii) show very young starburst ages of $10-50$ Myr. The median (16th, 84th percentiles) of $t_{\mathrm{SB}}$ are: 100 Myr (44, 154 Myr) for star-forming galaxies, 86 Myr (49, 187 Myr) for composites, 145 Myr (59, 366 Myr) for AGN hosts, and 384 Myr (369, 393 Myr) for weak-line galaxies. These ages are systematically younger than those inferred by converting PCA2 to $t_{\mathrm{PSB}}$ using the calibration of \citet{wild07}, though we note that this difference may be driven by different definitions of $t_{\mathrm{PSB}}$ versus $t_{\mathrm{SB}}$. 

What differentiates galaxies in quadrants (i) and (ii), and why are most composites and AGN hosts located in quadrant (ii)? Figure~\ref{f:Delta_MS_vs_PCA2_dust_and_IR} shows the $\Delta(\mathrm{MS})$--PCA2 plane, with galaxies color-coded by the dust optical depth towards the youngest stellar populations, $\tau_{V}$, derived using \texttt{Prospector} SED fits. Among galaxies above the star-forming main sequence, the galaxies in quadrant (ii), which are also Balmer-strong, are systematically more dust-obscured than the galaxies in quadrant (i). The median and median absolute deviation of $\tau_{V}$ for galaxies in quadrant (i) are 1.11 and 0.34 mag, while for quadrant (ii) they are 2.31 and 0.44 mag. Because far-infrared emission provides key constraints on young stellar populations and ongoing star formation in dusty systems, we verified that this trend is not driven by detection bias. As shown in the right panel of figure~\ref{f:Delta_MS_vs_PCA2_dust_and_IR}, the far-infrared detection fractions are comparable between the two quadrants.

\begin{figure}
	\centering
\includegraphics[width=\columnwidth]{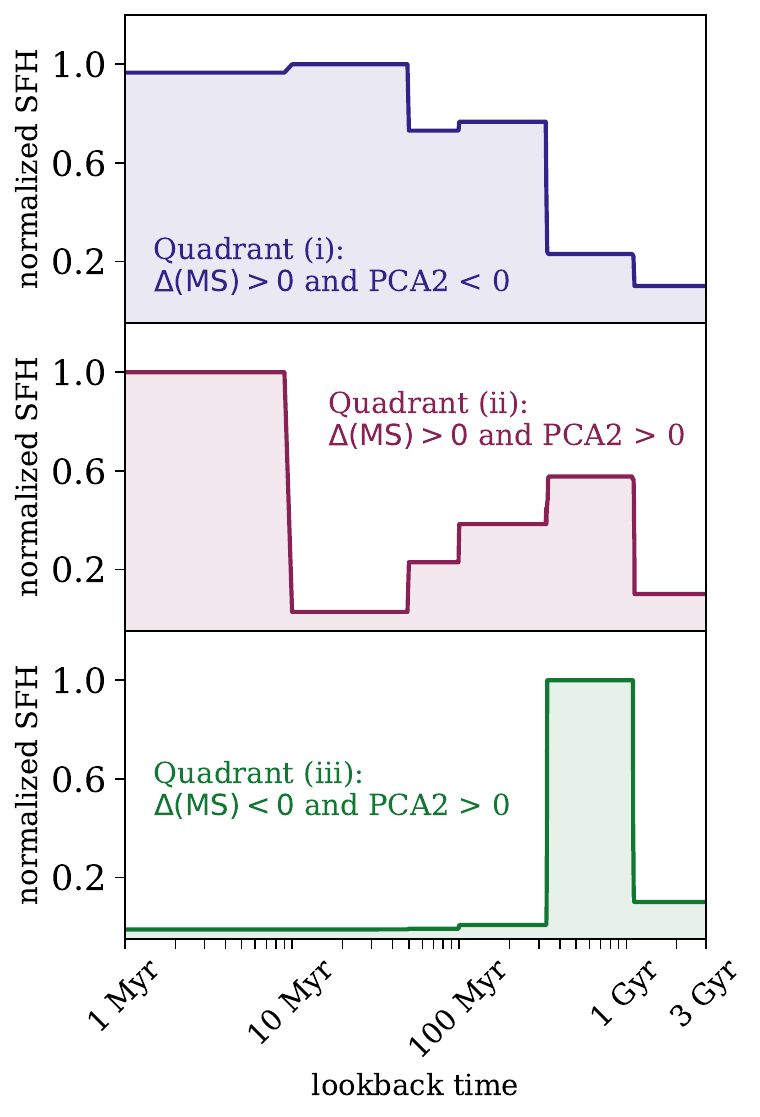}
\caption{\textbf{Characteristic SFHs of galaxies in different quadrants of the $\Delta(\mathrm{MS})$--PCA2 plane.} Each panel shows the median and rescaled SFH computed from all galaxies in a given $\Delta(\mathrm{MS})$--PCA2 quadrant. The top panel corresponds to quadrant (i), containing galaxies above the star-forming main sequence without a Balmer-strong continuum. The middle panel shows quadrant (ii), galaxies above the main sequence but exhibiting strong Balmer absorption lines. The bottom panel represents quadrant (iii), galaxies below the main sequence with strong Balmer absorption lines.}
\label{f:SFH_differences_between_classes}
\end{figure}

We find significant differences in the shapes of the \texttt{Prospector}-derived SFHs across the different quadrants. These trends have been verified by inspecting the SFHs of individual galaxies. In figure \ref{f:SFH_differences_between_classes}, we show representative SFH shapes for each quadrant. For this, we collect the derived SFHs in each quadrant and normalize them to an SFR of $1\,\mathrm{M_{\odot}\,yr^{-1}}$ at 3 Gyr. We then compute the median SFH per quadrant. To facilitate comparisons of SFH shapes, the figure shows the rescaled median SFH such that the SFR is $0.1\,\mathrm{M_{\odot}\,yr^{-1}}$ at 3 Gyr and $1\,\mathrm{M_{\odot}\,yr^{-1}}$ at its maximum. We also verified that the median SFH shapes are consistent across different galaxy classes (star-forming, composite, and AGN hosts) within each quadrant.

The combination of offset from the star-forming main sequence, the presence or absence of strong Balmer absorption lines, and the derived SFHs, together provide a self-consistent picture of the ways in which the \textit{Sparks} galaxies transition from starburst to post-starburst. Galaxies in quadrant (i) show low SFRs at early times, followed by a rise about 200 Myr ago, peaking 1–10 Myr ago. Their ongoing starburst is their first major starburst in the past 3 Gyr\footnote{With the current data, we cannot rule out the possibility that a weaker starburst occurred 300 Myr–1 Gyr ago but remains undetectable in the SFH because it is outshone by the ongoing starburst.}, explaining both their offset above the star-forming main sequence and the weak Balmer absorption lines. Galaxies in quadrant (ii) experienced their first significant starburst 100 Myr–1 Gyr ago, producing a post-burst population that gives rise to the strong Balmer absorption lines in their spectra. Their SFH rises again 1–10 Myr ago, representing a second starburst, which places them above the main sequence. Galaxies in quadrant (iii) underwent their most recent starburst 300 Myr–1 Gyr ago and have since ceased forming stars, making them Balmer-strong and positioned below the star-forming main sequence. We note, however, that the data do not necessarily imply that galaxies in quadrant (i) transition to post-starburst through quadrant (ii). These quadrants may represent distinct pathways in the transition from starburst to post-starburst (see e.g., \citealt{pawlik18}).

\begin{figure*}[p]
	\centering
\includegraphics[width=0.78\textwidth]{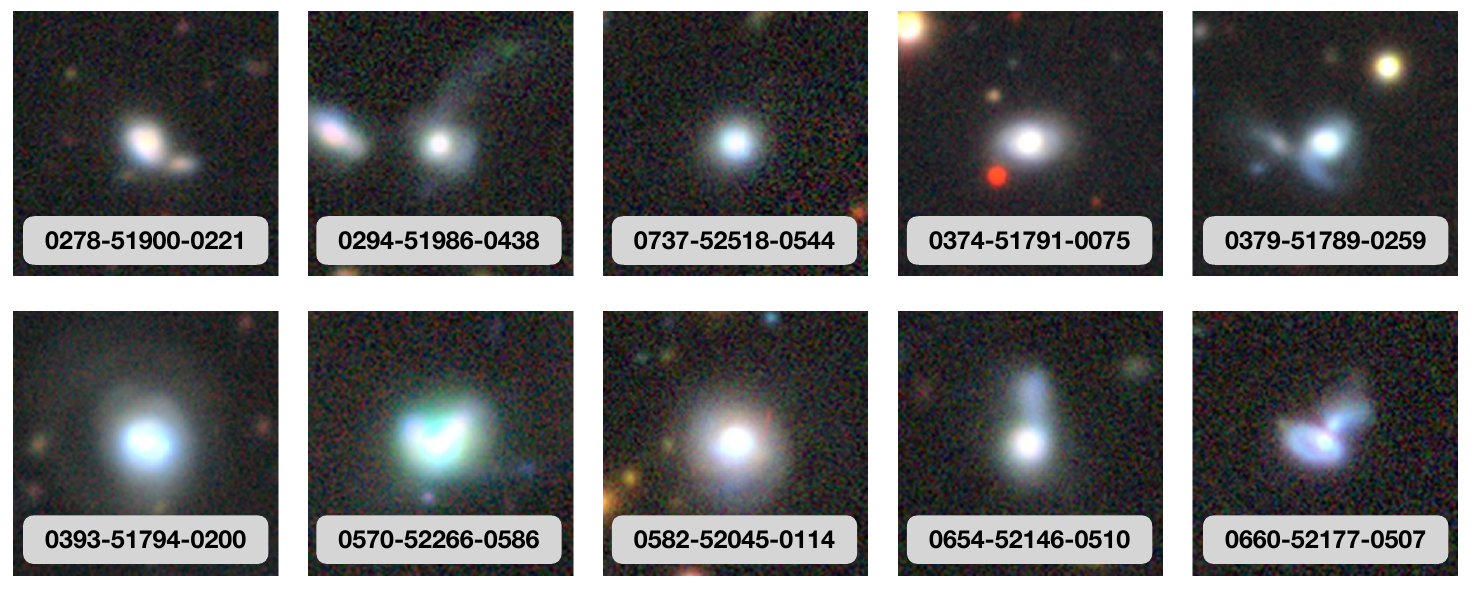}
\caption{\textbf{Image cutouts of galaxies undergoing their \textit{first} starburst: quadrant (i) in the $\Delta(\mathrm{MS})$--PCA2 plane.} Shown are $grz$-band cutouts from the \textit{Legacy Surveys Sky Viewer} (credit: Legacy Surveys / D. Lang; Perimeter Institute) obtained via the web interface: \url{https://yymao.github.io/decals-image-list-tool/}. Each cutout is 30\arcsec$\times$30\arcsec, with the SDSS identifiers (plate-MJD-fiberID) labeled. The galaxies in quadrant (i) often appear at an early stage of a major merger, many exhibiting blue colors and distinct companions.}
\label{f:normal_PCAMS_galaxies}
\end{figure*}

\begin{figure*}[p]
	\centering
\includegraphics[width=0.78\textwidth]{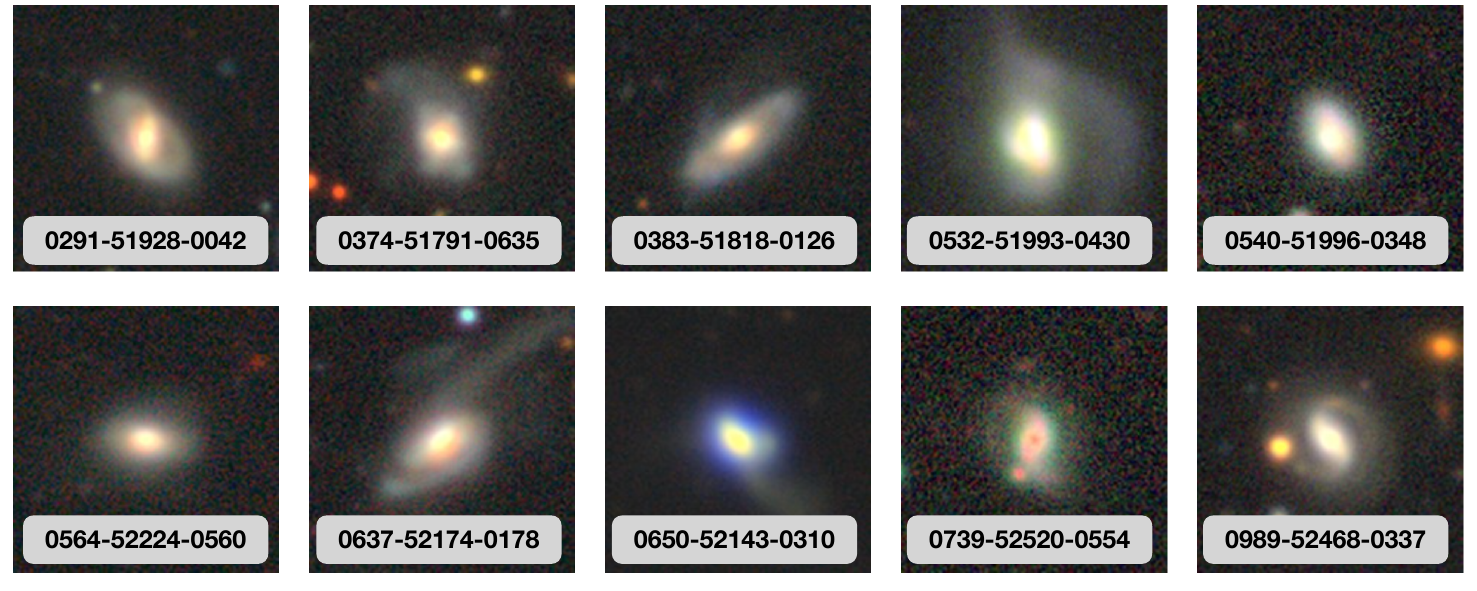}
\caption{\textbf{Image cutouts of galaxies undergoing their \textit{second} starburst, following a previous burst 100 Myr–1 Gyr ago: quadrant (ii) in the $\Delta(\mathrm{MS})$--PCA2 plane.} Similar to Figure~\ref{f:normal_PCAMS_galaxies}. Galaxies in quadrant (ii) often appear at a more advanced stage of merging, many displaying redder colors in their centers and bluer in the outskirts (driven by dust reddening), tidal features, and lacking clearly distinguishable companions.}
\label{f:wierd_PCAMS_galaxies}
\end{figure*}

\begin{figure*}[p]
	\centering
\includegraphics[width=0.78\textwidth]{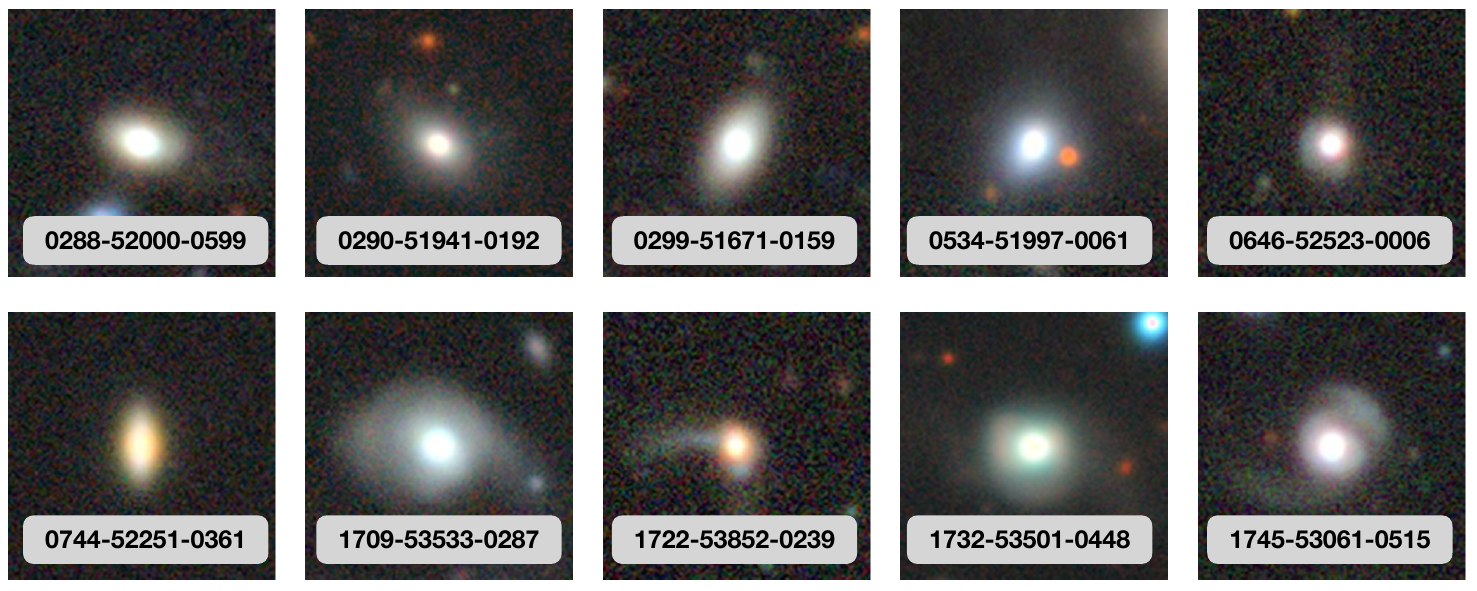}
\caption{\textbf{Image cutouts of galaxies that experienced their last starburst 300 Myr–1 Gyr ago: quadrant (iii) in the $\Delta(\mathrm{MS})$--PCA2 plane.} Similar to Figure~\ref{f:normal_PCAMS_galaxies}. Galaxies in quadrant (iii) typically appear post-merger, showing fewer tidal features, more prominent bulges, and redder colors throughout, consistent with aging stellar populations.}
\label{f:quenched_PCAMS_galaxies}
\end{figure*}

Inspection of galaxy images from the different quadrants supports this picture and places it in the broader context of galaxy evolution through major mergers. Galaxies in quadrant (i) often show blue colors and distinct companions, suggesting that their first ongoing starburst is triggered by an initial close passage between merging systems (Figure \ref{f:normal_PCAMS_galaxies}). Galaxies in quadrant (ii) no longer exhibit clear companions but instead display tidal features, indicating that they are close to or have recently undergone coalescence (Figure \ref{f:wierd_PCAMS_galaxies}). They also show pronounced color gradients, with redder central regions, consistent with higher dust optical depths, and bluer outskirts. Galaxies in quadrant (iii) show fewer tidal features and more prominent bulges (Figure \ref{f:quenched_PCAMS_galaxies}), with their colors appearing more uniformly red, consistent with aging stellar populations.

The striking over-representation of composite and AGN hosts in quadrant (ii) suggests that black hole accretion becomes visible through optical emission lines during the \textit{second, ongoing starburst}, and is much less evident during the first burst. The broad-band images, together with the estimated reddening, further suggest a connection to different merger stages: galaxies in quadrant (i) may correspond to an initial close passage, whereas those in quadrant (ii) may be approaching coalescence. Confirming this scenario will require a more quantitative assessment of merger stage from the imaging data, which is beyond the scope of this paper. This picture is consistent with the findings of \citet{ellison25}, who reported that the peak AGN excess occurs immediately after coalescence. It is also consistent with results from some hydrodynamical simulations of major mergers which find peak black hole accretion rate to SFR close to coalescence (e.g., \citealt{hopkins06, stickley14, capelo15, volonteri15}). These simulations suggest that black hole accretion can also peak at earlier stages of the merger, e.g., between first and second pericentre (\citealt{volonteri15}), but it is weaker and temporally de-correlated with the SFR. 

Examining the PCA plane proposed by \citet{wild07}, we confirm that this trend is a general property of SDSS local galaxies rather than a peculiarity of the \textit{Sparks} sample: composites and AGN hosts are predominantly located at PCA2 $>0$. It also unifies two results from the literature that might have been considered conflicting: (I) there is a delay between the starburst phase and the peak of nuclear optical AGN activity, of about 200 Myr \citep{wild10, yesuf14}, and (II) Balmer-strong galaxies with AGN signatures often show substantial far-infrared emission, placing them 0.3--1 dex above the star-forming main sequence \citep{baron22, baron23}. We suggest that the first result is driven by starburst ages inferred from SFHs constrained by optical spectroscopy or ultraviolet--optical photometry, which are primarily sensitive to the first burst that occurred 300 Myr--1 Gyr ago. In contrast, the far-infrared emission originates from the ongoing second burst triggered at coalescence. 

Our results suggest that AGN signatures are preferentially detected in actively starbursting systems, where the strong Balmer absorption lines serve as a fossil record of the galaxies’ earlier close passage that triggered the previous starburst. It remains possible that black hole accretion is also triggered during the first burst, but either produces weaker ionizing radiation compared to that emitted by the stars, or is heavily obscured by dust. The FIRE near-infrared spectra we collected for the \textit{Sparks} galaxies will allow us to test the latter scenario by probing gas emission in more dust-obscured regions.

\begin{figure}[t!]
	\centering
\includegraphics[width=1\columnwidth]{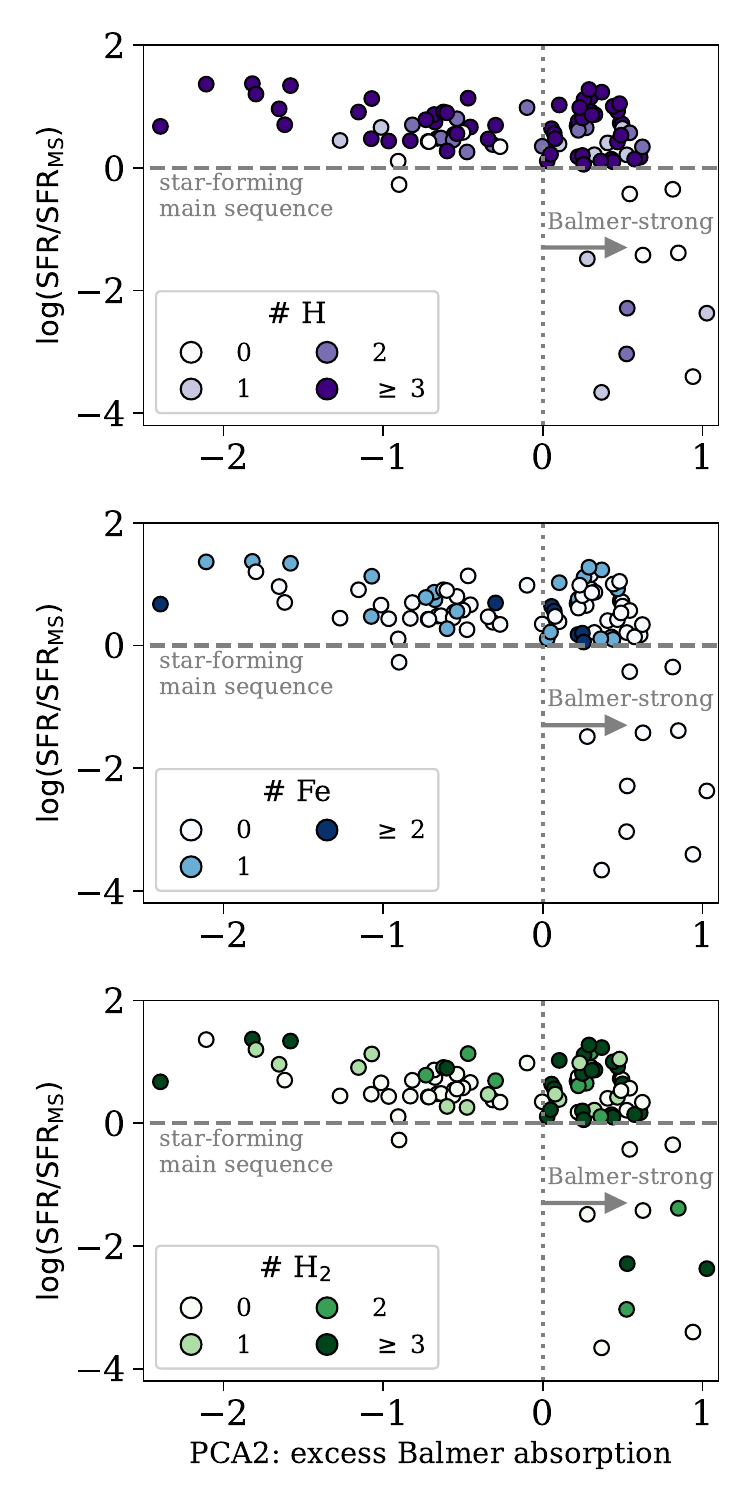}
\caption{\textbf{Detection of different near-infrared lines in the \textit{Sparks} galaxies.} Each panel shows the galaxies in the $\Delta(\mathrm{MS})$ versus PCA2 plane as described in section \ref{sec:results}. The horizontal dashed line marks the star-forming main sequence, and the vertical dotted line indicates the PCA2 threshold above which galaxies are classified as Balmer-strong. Each panel shows the number of detected hydrogen, iron, and H$_{2}$ lines in a FIRE spectrum. The lines we considered in the counting are as follows. Top panel: Pa$\gamma$, Pa$\beta$, Pa$\alpha$, Br$\delta$, and Br$\gamma$. Middle panel: $\text{Fe\,{\sc ii}}\lambda\, 12570{\mathrm{\AA}}$ and $\text{Fe\,{\sc ii}}\lambda\, 16436{\mathrm{\AA}}$. Bottom panel: H$_{2}$ 1-0 S(0) to S(4).}
\label{f:line_detection_vs_MS_and_PCA}
\end{figure}

\section{Survey science goals illustrated with FIRE data}\label{sec:science_goals}

The program is designed to study the properties of star formation, black hole accretion, and the multiphase gas and dust properties of galaxies transitioning from starburst to post-starburst. It focuses on massive galaxies ($M_{*} \sim 10^{10}$--$10^{11}\,\mathrm{M_{\odot}}$) at $z \sim 0.1$, and spans about three orders of magnitude in star formation rate ($\sim -2$ to $1$ dex with respect to the star-forming main sequence) and about 600 Myr in post-burst age. It includes star-forming galaxies, AGN, and galaxies with weak emission lines.

\begin{figure*}[t!]
	\centering
\includegraphics[width=1\textwidth]{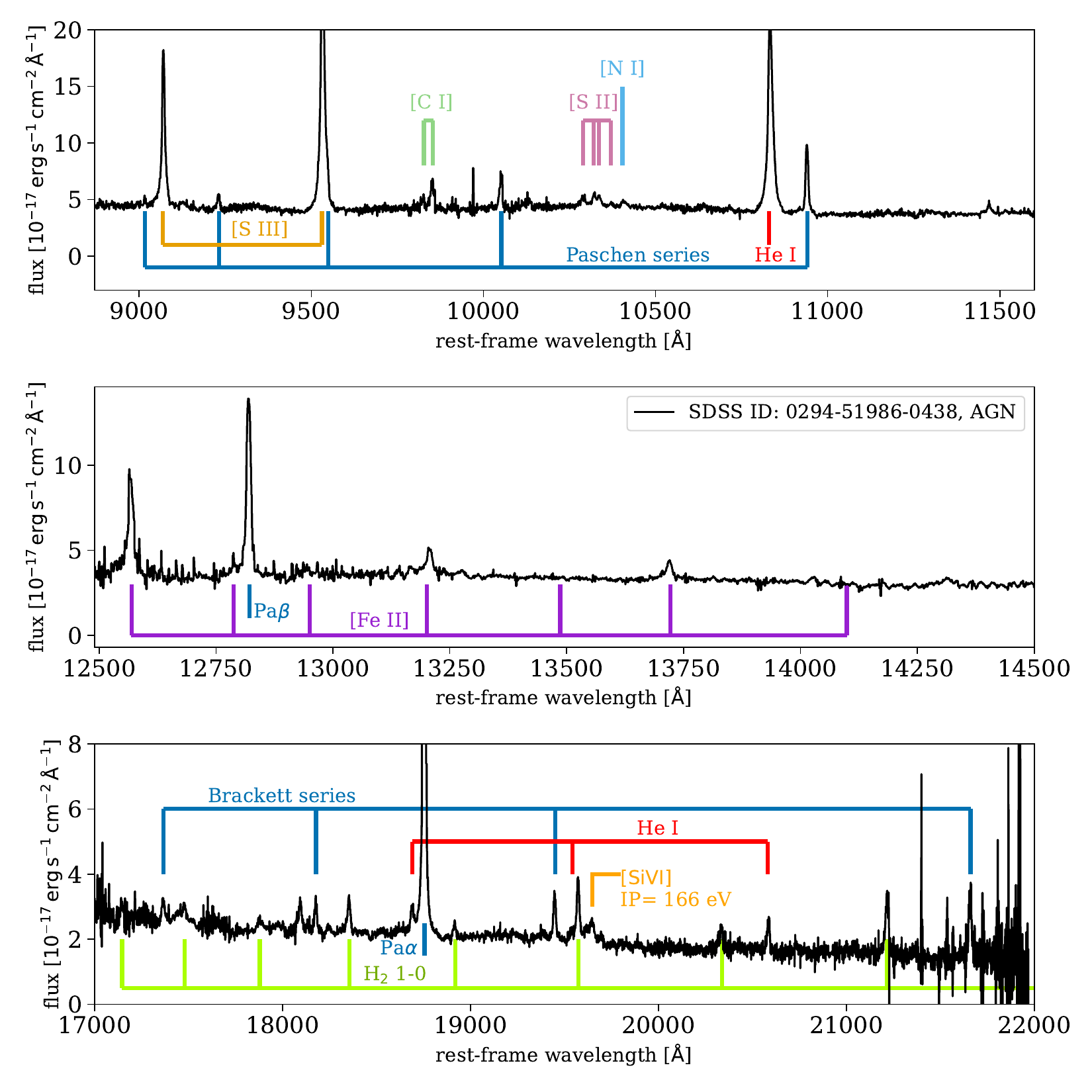}
\caption{\textbf{Example of a high SNR near-infrared spectrum of a galaxy classified as AGN.} The Magellan/FIRE spectrum of SDSS ID 0294-51986-0438, classified as AGN using standard line diagnostic diagrams. The spectrum is shown in air wavelengths and shifted to rest-frame using the SDSS redshift. The top panel shows a wavelength range within the $J$ band, the middle panel shows a range within $H$ band, and the bottom shows $K$ band. We detect more than 20 emission lines throughout the spectrum, and mark the line identifications in the different panels.}
\label{f:FIRE_spec_AGN_example}
\end{figure*}

The rest-frame near-infrared wavelength range probed by FIRE is ideal for this study, as it includes multiple hydrogen recombination lines, forbidden lines tracing warm ionized gas (e.g., $\text{[S\,{\sc iii}]}$, $\text{[Fe\,{\sc ii}]}$), multiple transitions of H$_{2}$ tracing hot molecular gas, and coronal lines tracing extremely high ionization gas ($\mathrm{IP > 100\, eV}$). The utility of our FIRE data is demonstrated in Figure~\ref{f:line_detection_vs_MS_and_PCA}, which shows high detection rates for hydrogen, iron, and H$_{2}$ lines across the full range of galaxy properties in our sample. Our higher SNR spectra, shown for example in Figure~\ref{f:FIRE_spec_AGN_example}, reveal more than 20 emission lines, proving rich diagnostics of multiphase gas. 

The survey is designed to serve as a local baseline for comparison with higher-redshift ($1 \lessapprox z \lessapprox 6$) starburst and post-starburst massive galaxies observed with JWST (e.g., \citealt{carnall23, carnall23b, dEugenio23, strait23, deGraaff24, setton24, slob24}). With the wealth of multi-wavelength information from far ultra-violet to far-infrared, it can be used to test various assumptions employed when studying such sources. In the sub-sections below, we describe our main science goals.

\subsection{Multi-wavelength census of star formation rate on different time scales}\label{sec:science_goals:SF}

A key goal is to measure the SFR by observing hydrogen recombination lines in the rest-frame near-infrared along the starburst to post-starburst sequence. Rest-frame optical and mid- to far-infrared wavelengths can paint conflicting pictures about the presence of star formation in Balmer-strong galaxies (e.g., \citealt{dressler83, dressler99, poggianti99, smail99, poggianti00, wild07, wild09, wild10, yesuf14, french15, french18, pawlik18, smercina18, wild20, baron22, smercina22, baron23, wu23, zhu25}; and see a review by \citealt{french21}). Being less susceptible to dust extinction, near-infrared hydrogen lines can probe obscured star formation on short timescales ($\sim$10 Myr).

The survey is designed to robustly measure the star formation rate by observing hydrogen recombination lines in rest-frame near-infrared along the starburst to post-starburst sequence. Being in rest-frame near-infrared, the hydrogen lines from the Paschen and Brackett series are less susceptible to dust extinction and can be used to probe obscured star formation. Combining the near-infrared hydrogen recombination lines with the mid- and far-infrared continuum emission, which traces dust-reprocessed light and thus star formation on longer timescales ($\sim$100 Myr; e.g., \citealt{kennicutt98, calzetti13}), the survey will allow us to study star formation on short ($\sim$10 Myr) and long ($\sim$100 Myr) timescales, and examine how they differ as the starburst ages and with respect to the star-forming main sequence. 

\begin{figure*}
	\centering
\includegraphics[width=1\textwidth]{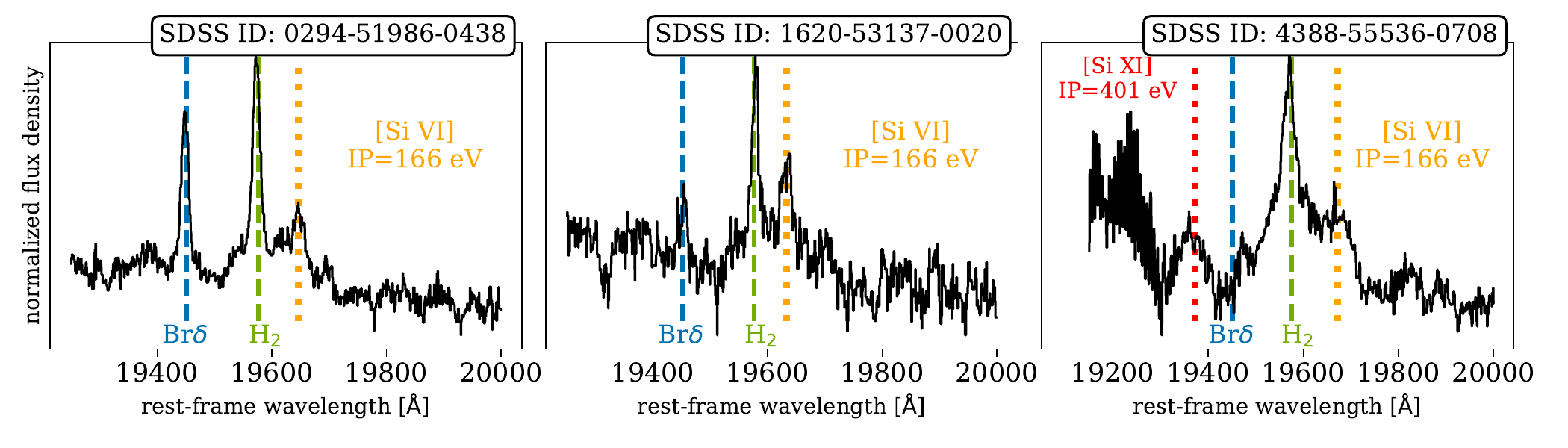}
\caption{\textbf{Detected coronal lines tracing highly-ionized gas.} The three panels show the rest-frame near-infrared spectra of three AGN in our sample where we detect the coronal line $\text{[Si\,{\sc vi}]}\lambda \, 1.96{\mathrm{\mu m}}$, tracing gas with IP of 166 eV, with high significance. The identified lines are redshifted by a few hundreds of $\mathrm{km \, s^{-1}}$ from systemic velocity. In SDSS ID 4388-55536-0708, we also identify $\text{[Si\,{\sc xi}]}\lambda \, 1.93{\mathrm{\mu m}}$, which traces gas with IP of 401 eV.}
\label{f:coronal_lines}
\end{figure*}

The comparison between the short and long timescale star formation rate tracers may provide additional constraints on the quenching timescale of the burst, a property that is often uncertain and degenerate when performing SPS modeling of multi-wavelength SEDs and/or optical spectra (e.g., \citealt{french18, wild20}). The comparison can also be used to test whether rest-frame optical and near-infrared observations are sufficient in identifying obscured starbursts, which is of particular importance for higher-redshift post-starburst galaxy samples observed with JWST, where rest-frame mid- and far-infrared observations are either unavailable or do not provide sufficiently stringent upper limits.

\subsection{A local baseline for rest-frame near-infrared line diagnostic diagrams}\label{sec:science_goals:nir_lines}

\begin{figure*}[t!]
	\centering
\includegraphics[width=1\textwidth]{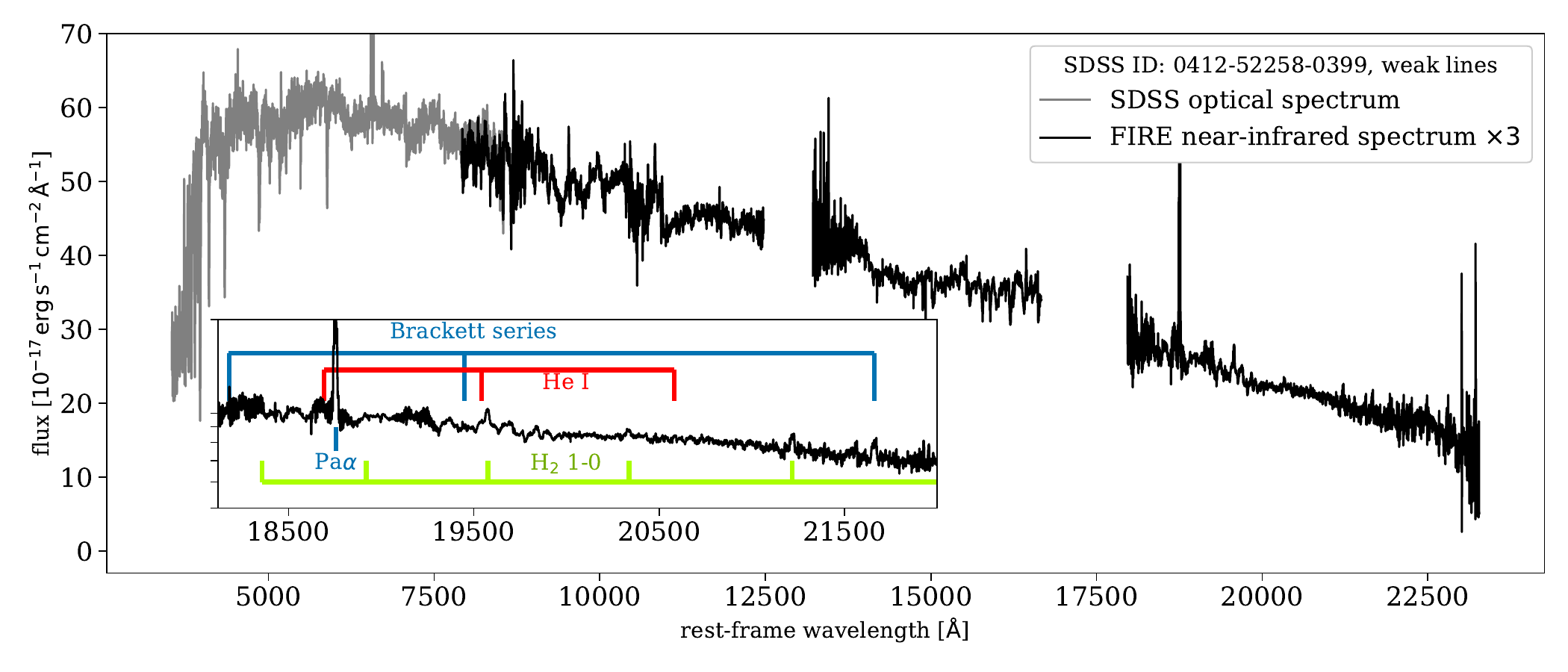}
\caption{\textbf{Example of a high SNR near-infrared spectrum of a galaxy with relatively weak emission lines.} Spectra of SDSS ID 0412-52258-0399, which is classified as having weak emission lines in the MPA-JHU catalog. Its H$\alpha$ EW is $\sim 9 \,\mathrm{\AA}$, so it does not pass the traditional criterion EW(H$\alpha$)$< 3\, \mathrm{\AA}$ for selection of E+A galaxies (e.g., \citealt{french21}). The gray line shows the SDSS optical spectrum, and the black line shows the Magellan/FIRE near-infrared spectrum, multiplied by 3. The inset in the bottom left shows a zoomed-in version of the $K$ band spectrum, where hydrogen, H$_{2}$, and possibly helium lines, are detected.}
\label{f:FIRE_spec_weaklines_example}
\end{figure*}

With the spectroscopic capabilities of JWST, rest-frame near-infrared line diagnostics are required to interpret spectra of high-redshift galaxies (e.g., \citealt{brinchmann23, calabro23, shapley24}). While several such diagrams have been suggested (e.g., \citealt{rodriguez-ardila04, riffel13, brinchmann23, calabro23} and references therein), they suffer from different uncertainties and are subject to different limitations, and their testing and validation have been limited by the smaller sample sizes of local galaxies observed in rest-frame near-infrared compared to the optical samples (see examples in \citealt{riffel06, davies14, mason15, lamperti17, riffel21}). For example, some diagrams rely on the ratio of multiphase gas tracers (e.g., H$_{2}$/Br$\gamma$) or lines from iron, which is heavily depleted onto dust grains (e.g., \citealt{osterbrock06, draine11}), complicating their interpretation.

Our survey is designed to provide a local baseline for testing and validation of different rest-frame near-infrared line diagnostic diagrams. It includes galaxies with different ionizing sources, and the spectral coverage in the rest-frame near-infrared probes all the relevant transitions. Combining the rest-frame optical and near-infrared spectral information, the survey will compare between the classification of the sources using optical and near-infrared diagnostics. Combining these with photoionization models, it will be possible to estimate the iron depletion factor in local galaxies along the starburst to post-starburst sequence, providing the first necessary step for applications of $\text{[Fe\,{\sc ii}]}$-based diagnostic diagrams at high redshifts. 

\subsection{Signatures of obscured black hole accretion in rest-frame near-infrared}

Supermassive black hole accretion is expected to be heavily obscured during the peak of a starburst (e.g., \citealt{sanders88, barnes96, genzel98, springel05a, hopkins06}). The rest-frame near-infrared may reveal the presence of an accreting black hole not seen in optical observations, providing local analogs to obscured AGN identified by JWST at high redshifts (e.g., \citealt{labbe23, yang23, maiolino24, suh24}). We aim to look for signatures of obscured accretion by (I) searching for broad kinematic components ($v \gtrapprox 1000 \, \mathrm{km\,s^{-1}}$) in near-infrared recombination lines without optical counterparts, and (II) searching for coronal lines tracing high ionization potentials (IP $\sim$130--450 eV; e.g., \citealt{lamperti17}).

These coronal lines, such as $\text{[Si\,{\sc vi}]}\lambda \, 1.96{\mathrm{\mu m}}$, are believed to be powered by photoionization by the hard ultraviolet and soft X-ray continuum produced by AGN. As shown in Figure \ref{f:coronal_lines}, we detect this line unambiguously in several AGN in our sample, and even higher-IP lines like $\text{[Si\,{\sc xi}]}\lambda \, 1.93{\mathrm{\mu m}}$ in one source. With these, the survey aims to map when, along the sequence, there is evidence for obscured black hole accretion.

\subsection{Hot molecular gas along the starburst to post-starburst sequence}\label{sec:science_goals:H2}

The near-infrared includes H$_{2}$ ro-vibrational transitions tracing hot molecular gas ($T \sim 1000-2000$ K; e.g., \citealt{dale05, davies_r14, emonts17}). The survey aims to examine the fraction of the hot molecular gas mass, temperature, and kinematics in systems evolving from starburst to post-starburst. While a sub-dominant phase in mass, hot molecular gas kinematics can trace the kinematics of the cold molecular gas reservoir (e.g. \citealt{muller_sanchez09, davies_r14}). Figure \ref{f:line_detection_vs_MS_and_PCA} shows that we detect H$_{2}$ in about half our sample. This includes detections in systems below the main sequence with post-burst ages of hundreds of Myrs (e.g., Figure \ref{f:FIRE_spec_weaklines_example}; see also \citealt{otter24}). 

\subsection{Dust properties along the starburst to post-starburst sequence}\label{sec:science_goals:dust}

The combined optical SDSS and near-infrared FIRE spectra result in a spectral coverage between $\sim$4000 \AA\xspace and $\sim$2.2 \mic\footnote{Two gaps are present in the infrared wavelength coverage due to strong telluric absorption between the $J$ and $H$ bands and between the $H$ and $K$ bands.}, with multiple detected hydrogen recombination lines from the Balmer, Paschen, and Brackett series. Assuming case-B recombination, and either a screen or a mixed gas-dust geometry (e.g., \citealt{calzetti01}), the observed fluxes of the different transitions can be used to place constraints on the dust reddening affecting the line-emitting gas. At the same time, optical to near-infrared continuum emission constrains the reddening of the stellar populations. By probing three orders of magnitude of star-formation with respect to the star-forming main sequence, the survey aims to constrain how dust reddening and geometry with respect to stars vary with respect to the star-forming main sequence.

In the violent environment of a starburst, there are multiple processes that may contribute to dust destruction (e.g., \citealt{draine79, mckee87, borkowski06, slavin15, dopita16}). The survey's near-infrared spectra show several $\text{[Fe\,{\sc ii}]}$ lines, tracing emission of iron in the gas phase. Since iron is heavily-depleted onto dust grains (e.g., \citealt{osterbrock06, draine11}), its flux strongly depends on the iron dust depletion factor, and may therefore be used to trace dust destruction processes. In section \ref{sec:science_goals:nir_lines} we discuss our aim to compare optical and near-infrared line diagnostic diagrams, where the latter use $\text{[Fe\,{\sc ii}]}$ lines. Using photoionization models with varying iron depletion factors to reproduce all the optical and near-infrared line ratios, we plan to examine the range of possible iron depletion factors that are consistent with our observations. Since $\text{[Fe\,{\sc ii}]}$ emission also depends on the gas excitation, it is necessary to include emission lines such as \siifull to better constrain the ionization sources. We plan to compare the derived iron depletion factors to those assumed in the literature or derived from observations (e.g., \citealt{pettini02, dopita16, calabro23, shapley24}), and to look for evidence of changes in the iron depletion factor across \textit{Sparks}.

\section{Summary}\label{sec:summary}

Understanding how galaxies evolve across cosmic time requires connecting their intense activity during the merger-triggered starburst phase to the subsequent quiescence observed in the local Universe. This transition is encapsulated in the post-starburst phase (E+A or Balmer-strong), a short-lived episode where black hole accretion emerges and shuts down, star formation decreases rapidly, and the molecular gas reservoirs are dramatically altered. There remain key open questions about the transition: (i) what is the true star formation rate of post-starburst galaxies, especially those with detected AGN, given the conflicting pictures painted by rest-frame optical and far-infrared tracers? (ii) when exactly does black hole accretion peak, and how does AGN activity impact the molecular gas and star formation? (iii) what is the timescale of molecular gas depletion, and what processes shape it? Discrepancies among observational diagnostics, especially where dust obscuration masks star formation or AGN activity, challenge our ability to disentangle quenching mechanisms.

To address these questions, we present \textit{Sparks}, a new infrared spectroscopic survey of galaxies transitioning from starburst to post-starburst at $z \sim 0.1$. The sample spans stellar masses of $10^{10}$–$10^{11}\,M_{\odot}$, covers nearly three orders of magnitude in star formation rate (from 1 dex above to 2 dex below the star-forming main sequence), and includes galaxies with diverse optical line ratios, classified as star-forming, composite, AGN, and weak-line systems. Using the FIRE echelle spectrograph on the Magellan Baade Telescope, we obtained 93 high-resolution near-infrared spectra (0.82–2.51 \mic) over 20 nights of observations. Combined with \textit{SDSS} optical spectroscopy, this yields nearly continuous coverage from rest-frame 0.4–2.2 \mic, supplemented by photometry spanning the far-ultraviolet to the far-infrared. The science goals of the survey are (section \ref{sec:science_goals}):\vspace{-0.1in}
\begin{enumerate}\setlength\itemsep{-0.4em}
    \item Building a multi-wavelength census of star formation rates on short ($\sim$10 Myr) and long ($\sim$100 Myr) timescales, by combining rest-frame optical spectroscopy, mid- and far-infrared dust emission, and near-infrared hydrogen recombination lines. 
    \item Establishing a local baseline for rest-frame near-infrared line diagnostics, and benchmarking them against the well-established optical line diagnostics. 
    \item Identifying obscured black hole accretion through rest-frame near-infrared recombination and coronal lines not accessible in the optical.
    \item Tracing hot molecular gas using ro-vibrational H$_{2}$ transitions, and linking it to the evolution of star formation and cold gas reservoirs.
    \item Constraining dust properties and geometry using multiple hydrogen line series and iron lines sensitive to dust depletion and destruction.
\end{enumerate}

This paper describes the sample selection (Section~\ref{sec:sample}) and observing program (Section~\ref{sec:fire}), including data reduction and slit loss corrections. In a companion paper (Baron et al., submitted), we present the collection of far-ultraviolet to far-infrared photometry and perform fitting of the multi-wavelength SEDs and optical spectroscopy, exploring different fitting codes, model assumptions, and input data. A key outcome of that analysis is the ability to obtain more reliable instantaneous star formation rate estimates for galaxies with optical line ratios inconsistent with pure star formation (composites and AGN), where the standard H$\alpha$ flux cannot be used.

With the physical properties derived from optical spectroscopy and multi-wavelength SED fitting, the \textit{Sparks} galaxies can be roughly divided into three groups, with the age of the $< 1$ Gyr stellar populations varying smoothly across them (Section~\ref{sec:results}):\vspace{-0.1in}
\begin{enumerate}\setlength\itemsep{-0.4em}
    \item Galaxies above the star-forming main sequence with weak Balmer absorption lines, whose star formation histories indicate that they are experiencing their first major burst within the past $\sim$3 Gyr (quadrant i). They are relatively less obscured by dust, and their images often reveal clear companions, consistent with an early merger stage where the starburst is triggered by a close passage. 
    \item Galaxies above the star-forming main sequence with strong Balmer absorption lines, revealing a substantial post-burst population (quadrant ii). Their star formation histories show an earlier burst 300 Myr–1 Gyr ago and a second, ongoing burst peaking within the past $\sim$10 Myr. They are more heavily reddened, with images showing tidal features but fewer companions, and redder centers with bluer outskirts--consistent with systems approaching or just after coalescence.  
    \item Galaxies below the star-forming main sequence with strong Balmer absorption lines, whose star formation histories suggest a burst 300 Myr–1 Gyr ago followed by little subsequent activity (quadrant iii). Their images show fewer tidal features, more prominent bulges, and uniformly red colors.  
\end{enumerate}

Most of the \textit{Sparks} composites and AGN hosts fall into the second group. This suggests that black hole accretion becomes visible through optical emission lines during the second, ongoing starburst. The imaging data further qualitatively indicate that these systems may be close to coalescence. Their Balmer-strong continua provide a fossil record of an earlier burst probably triggered at an earlier close passage. This picture unifies two results from the literature that have perhaps been viewed as conflicting: (i) optical spectroscopy and ultraviolet–optical photometry imply a $\sim$200 Myr delay between the starburst and the onset of black hole accretion \citep{wild10, yesuf14}, while (ii) many Balmer-strong galaxies with AGN show far-infrared luminosities placing them 0.3–1 dex above the main sequence \citep{baron22, baron23}. We suggest that (i) is based on features sensitive to the first, past starburst, whereas (ii) is based on features sensitive to the second, ongoing burst. This result is also in line with recent studies that used galaxy images to time the merger, finding a peak in AGN signatures at coalescence, and with results from numerical simulations.

The new near-infrared spectra will enable us to search for deeply obscured black hole accretion during the initial starburst phase, prior to coalescence. They will also allow us to pinpoint the onset of coronal lines from highly ionized gas (ionization potential $>$ 100 eV) and to track the evolution of multiphase gas and dust throughout the starburst to post-starburst sequence.

\section*{Data Availability}

We provide a catalog, \texttt{Sparks\_info}, containing observational and derived properties for all galaxies in the \textit{Sparks} survey. The catalog includes basic observational identifiers (\texttt{plate}, \texttt{MJD}, \texttt{fiberid}, and \texttt{specObjID}), as well as source coordinates (RA, DEC) and redshifts. We list the \citet{wild07} first two principal components (PCA1 and PCA2), and the emission-line classification from the BPT diagram.  

Stellar masses and star formation rates from the MPA--JHU catalog are included for both the fiber and total galaxy light, with associated 16th, 50th, and 84th percentile uncertainties. We also provide emission-line fluxes and uncertainties for \oiii, \hbeta, \nii, \halpha (including EW), and \sii.

In addition, we include results from \texttt{Prospector} fits used for our final adopted physical properties. These comprise instantaneous star formation rates, stellar masses, optical depths towards the stellar birth clouds ($\tau_V$), starbust ages ($t_{\mathrm{SB}}$), and total infrared luminosities ($L_{\mathrm{TIR}}$), each reported with median and 16th/84th percentile values. The full tables of best-fitting parameters with \texttt{Prospector} and \texttt{MAGPHYS} are available through Paper II (Baron et al. submitted). 

We further provide the \texttt{FIRE\_obs\_info} catalog, which includes the information listed in table \ref{tab:fire_obs} for all the \textit{Sparks} galaxies.

\acknowledgments{
We thank D.~French, K.~Rowlands, and V.~Wild for providing valuable feedback that improved the presentation and interpretation of the results.

We thank the referee for their careful reading and constructive feedback, and for their thoughtful consideration of both papers.

During the data collection and initial analysis, D.~Baron was supported by the Carnegie--Princeton Fellowship. She is grateful to the Carnegie Observatories staff for their support during survey planning, in particular I.~Thompson for Magellan telescope scheduling, and to D.~Kelson, D.~Newman and G.~Rudie for valuable advice on observing with the FIRE spectrograph. D.~Baron expresses her deepest gratitude to the Las Campanas Observatory staff for their continuous support during the FIRE observations: Carlos Contreras, Mat\'{\i}as D\'{\i}az, Gonzalo D\'{\i}az, Carla Fuentes, Hern\'an Nu\~nez, Hugo Rivera, Mauricio Mart\'{\i}nez, and Gabriel Prieto.}\vspace{1.8cm }

\software{Astropy \citep{astropy13, astropy18, astropy22},
	   IPython \citep{perez07},
          matplotlib \citep{hunter07},
          PypeIt \citep{prochaska20},
          scikit-learn \citep{pedregosa11}, 
          SciPy \citep{scipy01},
		  }\vspace{2cm}

\bibliography{ref.bib}

\end{document}